\documentclass[fleqn,usenatbib]{mnras}

\usepackage{newtxtext,newtxmath}
\usepackage[T1]{fontenc}
\DeclareRobustCommand{\VAN}[3]{#2}
\let\VANthebibliography\thebibliography
\def\thebibliography{\DeclareRobustCommand{\VAN}[3]{##3}\VANthebibliography}
\usepackage{graphicx}
\usepackage{subcaption}
\usepackage{amsmath}
\usepackage{siunitx}
\newcommand{\tiu}{\,\mathrm{J}\,\mathrm{m}^{-2}\,\mathrm{K}^{-1}\,\mathrm{s}^{-1/2}}
\title[Thermophysical Modelling of Binary Asteroids]{Thermophysical Modelling of Eclipse and Occultation Events in Binary Asteroid Systems}

\author[S.~L. Jackson and B. Rozitis]{
Samuel L. Jackson,\thanks{E-mail: samuel.jackson@open.ac.uk \& contact@samuelleejackson.co.uk}
and Benjamin Rozitis\\
School of Physical Sciences, The Open University, Milton Keynes, MK7 6AA, UK
}

\date{Accepted 2024 September 19. Received 2024 September 19; in original form 2024 April 19.}
\pubyear{2024}

\begin{document}
\label{firstpage}
\pagerange{\pageref{firstpage}--\pageref{lastpage}}
\maketitle

\begin{abstract}
Binary systems comprise approximately 15 per cent of the near-Earth asteroid population, yet thermal-infrared data are often interpreted for these bodies as if they are single objects.
Thermal-IR light curves of binary asteroids (3905) Doppler and (175706) 1996 FG3 are analysed using an adaptation of the Advanced Thermophysical Model, deriving new constraints on their thermal inertias as $\Gamma = 114 \pm 31\,\tiu$ and $\Gamma = 142 \pm 6\,\tiu$, respectively.
We determine that this adapted model is suitable for binary systems where their primary rotation to secondary orbit period ratios can be approximately characterised by small integers.
Objects with more complex orbital states require a model with alternative temperature convergence methodologies.
Thermal inertia is shown to have a strong effect on binary thermophysical light curve morphology, introducing significant modulations both inside and outside of mutual event times.
The depth of eclipse events are shown to be suppressed at longer wavelengths due to the sensitivity to cooler parts of the surface, meanwhile surface roughness is shown to have little effect on the thermal light curve morphology.
A proof of concept model for the (65803) Didymos system is demonstrated, showing how such a binary model could be used to study the system during the European Space Agency's Hera mission, and the applicability of this adapted model to NASA's Lucy mission is also briefly discussed.
\end{abstract}

\begin{keywords}
minor planets, asteroids: general -- radiation mechanisms: thermal -- infrared: planetary systems
\end{keywords}

\section{Introduction}

    Due to the primitive nature of asteroids, studying their properties allows us to constrain the conditions present in the pre-solar nebula.
    The asteroids have been altered significantly over billions of years through space-weathering, gravitational interactions with planets, impacts, and thermal processes.
    Therefore, to trace back the properties of the objects we see today we must understand the physical and dynamical processes that have acted and continue to act on them.

    The proportion of binaries among the near-Earth asteroid (NEA) population is estimated to be $15\pm4\%$ \citep[for binaries with primary to secondary diameter ratios $\geq 0.18$, $D_P > \SI{0.3}{\kilo\metre}$;][]{2006Icar..181...63P}.
    Understanding binary formation and evolution is therefore critical to our understanding of the wider NEA population and beyond.
    Binary asteroid systems are thought to form through impacts \citep{2004Icar..167..382D}, tidal effects during close planetary encounters \citep[e.g.,][]{1996Natur.381...51B,1998Icar..134...47R,2006Icar..180..201W}, and spin-up and subsequent rotational fission through the Yarkovsky-O'Keefe-Radzievskii-Paddack (YORP) effect \citep{2000Icar..148....2R,2008Natur.454..188W}.
    Studying the thermal emission of binary asteroid systems can allow us to determine regolith properties and constrain the effect of thermal recoil forces such as the YORP and the Binary-YORP effects \citep[BYORP;][]{2005Icar..176..418C}.

    The study of the thermal emission from asteroid surfaces is achieved through the use of thermal models \citep{2015aste.book..107D}.
    The simplest thermal models assume a spherical shape for the asteroid, and vary from assuming instantaneous equilibrium between absorbed solar radiation and the radiation emitted from the surface of an asteroid in the Standard Thermal Model \citep[STM;][]{1986Icar...68..239L}, to assuming equal thermal emission around the entire body in the Fast Rotating Model \citep[FRM;][]{1989aste.conf..128L}.
    The Near-Earth Asteroid Thermal Model \citep[NEATM;][]{1998Icar..131..291H} includes a thermal-infrared beaming free parameter which allows for the effects of surface roughness and non-zero thermal inertia.
    Thermal inertia describes the resistance to temperature change of a material, and can be used to gain an estimate of regolith particle size \citep{2013Icar..223..479G} or the porosity of rocks and boulders on the surface \citep{2019NatAs...3..971G,2021NatAs...5..766S}.
    The thermal inertia is defined as $\Gamma = \sqrt{k \rho C_\text{P}}$, where $k$ is the thermal conductivity, $\rho$ is density, and $C_\text{P}$ is the specific heat capacity.

    Thermophysical models, however, provide a more complete understanding of the thermal emission from asteroid surfaces.
    These models use shape models of the asteroids instead of spherical shapes to calculate more accurate surface and sub-surface temperature distributions.
    These temperature distribution calculations account for the specific geometry and orientation of the asteroid, and can be calculated for a range of regolith properties such as thermal inertia and surface roughness.
    The inclusion of surface roughness is motivated by the observation of non-Lambertian thermal emission from the Lunar surface \citep[also known as the thermal-IR beaming effect;][]{1930ApJ....71..102P}.
    Surface roughness is typically included in thermophysical modelling through fractional coverage of craters, which has been shown to successfully reproduce this thermal-IR beaming effect \citep{1968JGR....73.5281B,1971Moon....3..189S,1971Moon....2..279W}.
    The measurement of surface roughness indicates the presence of uneven surfaces at scales similar to or larger than the thermal skin depth, defined as $d_S = \sqrt{kP/\pi\rho C_\text{P}}$, where $k$ is the thermal conductivity, $P$ is the rotation period of the asteroid, $\rho$ is the density, and $C_\text{P}$ is the specific heat capacity.

    Thermal inertias have been measured for many asteroids through thermophysical modelling, although constraints on thermal inertias of binary near-Earth asteroids have only been determined in detail (i.e. over a few geometries to remove the degeneracy with surface roughness, allowing surface roughness to be determined as a free parameter) using a thermophysical model for three systems in the literature; (1862) Apollo \citep[$140^{+140}_{-100}\tiu$;][]{2013A&A...555A..20R}, (276049) 2002 CE26 \citep[$170 \pm 30\tiu$;][]{2018MNRAS.477.1782R}, and (65803) Didymos \citep[$320 \pm 70\tiu$;][]{2024PSJ.....5...66R}.
    The value obtained for (65803) Didymos is comparable to those derived by \citet{2023PSJ.....4..214R} of $\Gamma = 260 \pm 30\,\tiu$ and $\Gamma = 290 \pm 50\,\tiu$ from MIRI and NIRSpec observations with JWST, respectively.
    A thermal inertia value has been derived for the binary asteroid (175706) 1996 FG3 as $120 \pm 50\tiu$ \cite{2011MNRAS.418.1246W}, although the observations only cover a single observation geometry and therefore the uncertainties on the derived thermal inertia are compounded by the lack of constraint on the surface roughness.
    Their work also modelled the components individually and combined their fluxes without accounting for mutual events.
    A thermal inertia of $80 \pm 40\,\tiu$ was subsequently derived by \citet{2014MNRAS.439.3357Y}, with the observations modelled covering two different geometries allowing for a better constraint on the roughness, although their work modelled (175706) 1996 FG3 as a single body.
    
    A range of main-belt binary asteroids had thermal inertias calculated by \citet{2012Icar..221.1130M}, with the values typically found to be $\leqslant\sim 100\tiu$.
    Their work, however, did not allow the roughness fraction to vary freely.
    Instead, a range of roughness fractions and crater opening angles were sampled and the range of thermal inertias reported as the range of the minimum-$\chi^2$ thermal inertias across the sampled roughness models.
    \citet{2011Icar..212..138D} calculate an average thermal inertia of a sample of binary near-Earth asteroids as $\Gamma = 480 \pm 70\tiu$ using the NEATM \citep{1998Icar..131..291H} and converting the fitted beaming parameters to thermal inertias through comparison with simulations from a thermophysical model that have been fitted with the NEATM.
    This average thermal inertia contrasts significantly with an average thermal inertia of $\Gamma = 200 \pm 40\tiu$ for non-binary NEAs measured in the same way \citep{2007Icar..190..236D}.
    The higher thermal inertia suggests an absence of fine regolith on the surfaces of binary asteroids, perhaps as a result of binary formation processes such as YORP-driven spin-up and subsequent rotational fission \citep{2008Natur.454..188W}.
    A common thread across all of these studies is that all have either modelled the binary system as a single object or treated the two components independently and therefore did not model eclipse and occultation events in the thermal light curves.
    \citet{2010Icar..205..505M} modelled the binary Trojan asteroid (617) Patroclus during eclipse events using a binary thermophysical model that accounted for shadowing of one component of the binary onto the other for fully synchronous systems.
    The fully-synchronous nature of the (617) Patroclus system allowed for a single-body TPM to be used, with the input shape containing two disjoint components, with a thermal inertia derived as $\Gamma = 20 \pm 15\tiu$.

    In this work, we have developed a binary TPM that works not only for fully synchronous rotation, but also for objects where the secondaries orbit at different rates to the rotation of the primary.
    However, this is limited to objects where there is an integer ratio between the orbital period of the secondary and the rotation period of the primary, and assuming the secondary rotation is synchronous with its orbit (we will refer to these types of systems as semi-synchronous in this manuscript).
    In Section~\ref{sec:TPM}, we outline the basis of the binary TPM, and discuss some of the assumptions made to simplify the model.
    The binary asteroids studied in this paper, along with the data used to model each of them are outlined in Section~\ref{sec:TPM}.
    In Section~\ref{sec:fitting}, we derive thermophysical properties for the targets using our binary TPM.
    The sensitivity of binary thermal infrared light curves and their mutual events to different parameters such as thermal inertia, roughness, and wavelength is discussed in Section~\ref{sec:sensitivity}.
    In Section~\ref{sec:discussion}, we outline use-cases of this binary TPM for interpretation of data from ESA's Hera mission and other opportunities for its application such as for NASA's Lucy mission to the Jupiter Trojans.

\section{Thermophysical Modelling}\label{sec:TPM}

    \subsection{Binary TPM}\label{sec:model}

        The binary thermophysical model used in this work is an adaptation of the Advanced Thermophysical Model \citep[ATPM;][]{2011MNRAS.415.2042R}.
        The model has been verified to reproduce the thermal-IR beaming effect discussed previously when assuming properties measured in-situ during the \textit{Apollo} missions \citep{2011MNRAS.415.2042R}.
        The ATPM is a single-object thermophysical model that has been widely used to interpret both disk-integrated \citep[e.g.,][]{2011MNRAS.418.1246W,2014A&A...562A..48L,2014A&A...568A..43R,2014Natur.512..174R,2017MNRAS.464..915R,2019A&A...627A.172R,2019A&A...631A.149R,2019NatAs...3..341D,2021MNRAS.507.4914Z,2022MNRAS.515.4551R,2023MNRAS.525.4581D,2023PSJ.....4..214R} and disk-resolved \citep[e.g.,][]{2020SciA....6.3699R,2020JGRE..12506323R,2022JGRE..12707153R,2024PSJ.....5...92R} thermal-IR data of asteroids.

        For full details on the ATPM, the reader is directed to \citep{2011MNRAS.415.2042R,2012MNRAS.423..367R,2013MNRAS.433..603R}, and the principal details of the model are reproduced here.
        The ATPM works by calculating the surface (and sub-surface) temperature distribution, and subsequent thermal emission, of an asteroid given a set of input parameters such as Bond albedo, thermal inertia, asteroid shape and spin-state.
        The one-dimensional heat diffusion equation (Equation~\ref{eq:heatDiff}) is solved for each facet of the asteroid shape model as the asteroid rotates, making use of two boundary conditions: a surface boundary condition (Equation~\ref{eq:surfaceBoundary}) to balance incoming energy from the Sun and scattered sunlight from interfacing shape facets with emitted energy through thermal emission and heat conduction into the sub-surface, and a boundary condition that ensures that the amplitude of temperature variations tends to zero with increasing depth (Equation~\ref{eq:adiabatic}).
        \begin{equation}\label{eq:heatDiff}
            \frac{dT}{dt} = \frac{k}{\rho C_\text{P}}\frac{d^2 T}{dx^2}
        \end{equation}
        \begin{equation}\label{eq:surfaceBoundary}
        \begin{split}
            \left( 1 - A_\text{B} \right) \left( \left[ 1 - S(t) \right]\psi(t)F_\text{SUN} + F_\text{SCAT}\right) + F_\text{RAD}& +\\ k\left( \frac{dT}{dx} \right)_{x=0}
            - \varepsilon\sigma T^{4}_{x=0} = 0&
        \end{split}
        \end{equation}
        \begin{equation}\label{eq:adiabatic}
            \left( \frac{\partial T}{\partial x} \right)_{x\rightarrow\infty} \rightarrow 0
        \end{equation}

        In Equation~\ref{eq:heatDiff}, the parameters are defined as follows: $T$ is temperature, $t$ is time, $k$ is thermal conductivity, $\rho$ is density, $C_\text{P}$ is the specific heat capacity, and $x$ is depth.
        In Equation~\ref{eq:surfaceBoundary}, the parameters are defined as follows: $A_\text{B}$ is the Bond albedo, $S(t)$ calculates the shadowing of each facet, $\psi(t)$ represents the cosine of the illumination angle for each facet, $F_\text{SUN}$ is the integrated solar flux at the distance of the asteroid from the Sun, $F_\text{SCAT}$ represents multiple-scattering of incident sunlight, $F_\text{RAD}$ represents self-heating from interfacing shape facets, $\varepsilon$ is the bolometric emissivity, and $\sigma$ is the Stefan-Boltzmann constant.

        Surface roughness is modelled using a fractional coverage of hemispherical craters, which themselves comprise 100 sub-facets.
        The temperatures of the crater sub-facets are calculated by solving the heat diffusion equation with the same boundary conditions as the shape facets as detailed previously.

        Once the temperatures are calculated for a given thermal inertia, the thermal flux emitted by each facet is calculated using the Planck function.
        This flux is then modified to account for spectral emissivity, facet area, distance to the observer, and emission angle to calculate the flux received by the observer, which is summed across all visible facets.
        This is then repeated for the roughness facets, providing two flux estimates: one for the `fully smooth' case, and one for the `fully rough' case.
        The flux for a given trial roughness fraction, $f_R$, is then a linear combination of the smooth and rough fluxes according to this fraction.
        The roughness fraction can be interpreted in terms of surface RMS slope by the following conversion: $\text{RMS Slope} = 49\sqrt{f_R}$.
        The flux can then also be scaled for a varying diameter if the diameter is not known prior to fitting.

        The fluxes are calculated for a range of trial thermal inertia values, and chi-squared minimisation is used to find the diameter, roughness fraction, and thermal inertia combinations that match the observed thermal-infrared fluxes.
        Acceptable parameter values are defined at the 3-sigma level as any combination that provided a $\chi^2$ value within $\Delta\chi^2 = 14.2$ of the minimum $\chi^2$ value for 3 free parameters and within $\Delta\chi^2 = 16.3$ of the minimum $\chi^2$ value for 4 free parameters \citep[p. 697;][]{2002nrca.book.....P}.
        The fitted parameter values and their uncertainties are then taken as the mean and standard deviation of the acceptable parameters.

        Thermophysical modelling of binary asteroids poses some unique challenges depending on the state of the system.
        A simple binary system that is fully synchronous can be treated simply as a single body in the code, with the input shape model consisting of two disjointed components, as was the case for the modelling of (617) Patroclus by \citet{2010Icar..205..505M}.
        A non-synchronous system can have their components rotated/orbited at different velocities and the temperatures calculated at each step.
        However, binary systems that are not synchronous are challenging to model.
        The surface temperatures calculated using Equations~\ref{eq:heatDiff}-\ref{eq:adiabatic} are iterated over many rotations of the shape until the facet temperatures converge to within $10^{-3}\si\kelvin$ between successive iterations.
        If a secondary in a binary system is orbiting non-synchronously with the rotation of the primary, then an eclipse shadow may not be cast onto the same part of the primary's surface after one orbit of the secondary.
        This means that we can not simply keep rotating the system until the temperatures converge, as surface elements may have different surface temperatures from one rotation to the next depending on where the secondary is in its orbit.
        The only way to ensure convergence using the currently implemented algorithms in the ATPM is by modelling systems that are semi-synchronous.
        Alternative approaches for temperature convergence in binary thermophysical models are discussed in Section~\ref{sec:future}.
        For the modelling of (175706) 1996 FG3 within this paper, the mutual orbit model has the primary rotating approximately 9 times for every 2 orbits of the secondary.
        Under this scenario, we can force the model into a spin-orbit resonance which allows for the model geometries to return to the exact same state after two orbits of the secondary, permitting temperature convergence over many iterations/revolutions of the model using the existing algorithms implemented in the ATPM.
        This is discussed further in Section~\ref{sec:inputs}.

        Thermophysical modelling of binary asteroids does not only pose challenges, but also the opportunity to evaluate the dependence of thermal properties on depth within asteroid regolith (which can't necessarily be assumed to have the same dependence as lunar regolith).
        \citet{2010Icar..205..505M} discuss that determining the thermal inertia of an asteroid using data within an eclipse event probes the effective thermal inertia at shallower depth scales compared to disk-averaged observations across an entire rotation at many geometries (which provides a measure of the effective thermal inertia across all relevant depth scales), due to the shorter duration of these eclipse events.
        As such, using binary thermophysical models such as that presented in this work and that of \citet{2010Icar..205..505M} may provide a method to probe the depth dependence of thermophysical properties by obtaining measurements of both the `diurnal thermal inertia' and the `eclipse thermal inertia' if the data are of sufficient quality.

    \subsection{Data Summary}

        To test our binary adaptation of the ATPM, we searched for objects with potential eclipse or occultation events in their thermal-IR light curves.
        From this we chose to model the synchronous binary system (3905) Doppler as a test of the simple fully-synchronous case, and (175706) 1996 FG3 as a test with a system that is approximately semi-synchronous.
        \textit{WISE} observations of these two targets were selected from the WISE All-Sky Single Exposure (L1b) Source Table for cryogenic observations, and from the NEOWISE-R Single Exposure (L1b) Source Table for non-cryogenic observations using the NASA/IPAC Infrared Science Archive\footnote{\url{https://irsa.ipac.caltech.edu/cgi-bin/Gator/nph-scan?mission=irsa&submit=Select&projshort=WISE}}.
        The observations were filtered to those that matched the expected asteroid position to within 1 arcsec and had a detection above the $3\sigma$ level.
        The \textit{WISE} magnitudes were then converted to fluxes, taking into account the red-blue calibrator discrepancy \citep{2010AJ....140.1868W}.
        An additional uncertainty was added in quadrature to the reported uncertainties of 2.8 per cent, 4.5 per cent, and 5.7 per cent in the W2, W3, and W4 bands respectively, based on comparisons of \textit{WISE} and \textit{Spitzer} observations \citep{2011ApJ...735..112J}.
        Colour corrections for sources that have different spectra to that of Vega \citep{2010AJ....140.1868W} were made to the model fluxes instead of the observed fluxes.
        The details of the \textit{WISE} observations used in this study are available in Table~\ref{tab:obsDetails}.

        \begin{table*}
            \centering
            \caption{Details of \textit{WISE} observations of (3905) Doppler and (175706) 1996 FG3 used in this work.}
            \label{tab:obsDetails}
            \begin{tabular}{lcccccc}
                \hline
                & Observation & & Heliocentric & Observer-centric & & \\
                Asteroid & Date [UT] & Survey & distance [au] & distance [au] & Phase angle [$\si\degree$] & Data \\ \hline
                (3905) Doppler        & 2010 March 06     & WISE Cryogenic            & 1.978 & 1.632 & 29.986    & 18 $\times$ W3 \\
                &&&&&& 17 $\times$ W4 \\
                (175706) 1996 FG3   & 2010 April 30     & WISE Cryogenic            & 1.220 & 0.611 & 55.390    & 16 $\times$ W2  \\
                &&&&&& 16 $\times$ W3 \\
                &&&&&& 16 $\times$ W4 \\
                                    & 2010 November 19  & NEOWISE Post-cryogenic    & 1.150 & 0.595 & 59.295    & 14 $\times$ W2  \\
                                    & 2021 February 03  & NEOWISE Reactivation      & 1.302 & 0.886 & 49.157    & 5 $\times$ W2  \\
                                    & 2021 February 15  & NEOWISE Reactivation      & 1.259 & 0.773 & 51.681    & 40 $\times$ W2  \\
                                    & 2021 March 08     & NEOWISE Reactivation      & 1.160 & 0.584 & 58.835    & 26 $\times$ W2  \\
                \hline
            \end{tabular}
        \end{table*}

    \subsection{Model Input Parameters}\label{sec:inputs}

        \begin{table}
            \centering
            \caption{Input parameters into the ATPM for (3905) Doppler and (175706) 1996 FG3.}
            \label{tab:modelInputs}
            \begin{tabular}{lcc}
                \hline
                Parameter & (3905) Doppler & (175706) 1996 FG3 \\ \hline
                $\lambda$ & $215\si\degree\,^{(1)}$ & $266\si\degree\,^{(2)}$ \\
                $\beta$ & $65\si\degree\,^{(1)}$ & $-83\si\degree\,^{(2)}$ \\
                $P_\text{prim. rot.}$ & $50.826\,\si\hour\,^{(1)}$ & $3.5890\,\si\hour\,^{(3)}$ \\
                $P_\text{sec. orb.}$ & $50.826\,\si\hour\,^{(1)}$ & $16.15076\,\si\hour\,^{(3)}$\\
                $A_\text{B}$ & $0.137\,^{(4)}$ & $0.01\,^{(5)}$\\
                $H$ & $12.657\,^{(4)}$ & $17.833\,^{(5)}$ \\
                $G$ & $0.215\,^{(4)}$ & $-0.041\,^{(5)}$ \\
                \hline \\
                \multicolumn{3}{p{\linewidth}}{\textbf{References.} (1) \citet{2020Icar..34513726D}, (2) \citet{2015Icar..245...56S}, (3) adapted from \citet{2015Icar..245...56S} value, (4) this work, (5) \citet{2011MNRAS.418.1246W}.}
            \end{tabular}
        \end{table}
        
        The input binary shape and spin/orbit solution used in this work for (3905) Doppler is that of \citet{2020Icar..34513726D}, which has the two bodies rotating synchronously with their mutual orbit.
        The mutual orbit period is set as $P=50.826\,\si\hour$, with a nominal mutual orbit pole of $\lambda = 215 \pm 2 \si\degree$ and $\beta = 65 \pm 2 \si\degree$.
        The input binary shape and spin/orbit solution used in this work for (175706) 1996 FG3 is that of \citet{2015Icar..245...56S}, which importantly does not have the two bodies rotating synchronously with their mutual orbit as with (3905) Doppler.
        The pole solution from \citet{2015Icar..245...56S} is $\lambda = 266 \pm 4\si\degree$ and $\beta = -83 \pm 4\si\degree$, and the orbital period of the secondary is $P_\text{orb.} = 16.1508\,\si\hour$.
        The rotation period of the primary from \citet{2015Icar..245...56S} is $P_\text{prim.} = 3.595195\,\si\hour$.
        These periods provide an approximate 9:2 spin-orbit resonance for this binary system (8.98:2 in reality), and so in this work we have kept the secondary orbit period fixed at the literature value (with a slight modification discussed in Section~\ref{subsec:FG3}) and set the primary rotation period to $P=3.589\,\si\hour$ to enable us to model the system as a semi-synchronous system as discussed in Section~\ref{sec:model}.
        This is only a 0.17 per cent change in the primary rotation period and so will have a negligible effect on the temperature distribution of the primary.
        The negligible effect on the temperature distribution of the primary by this change is justified through evaluation of the `dimensionless thermal parameter' $\Theta = \frac{\Gamma\sqrt{\omega}}{\varepsilon\sigma T_{SS}^4}$ \citep{1989Icar...78..337S}.
        This change in rotation period of 0.17 per cent will have only a 0.08 per cent effect on this dimensionless thermal parameter, which describes the surface temperature distribution of an atmosphereless body.
        The thermal-IR light curves from \textit{WISE} are dominated by the orbit of the secondary and the mutual events from the orbit of the secondary, and the primary has no rotational signal in the light curve as the primary shape is modelled as rotationally symmetric.
        This means that the modification of the primary rotation period has no effect on the thermal-IR light curve morphology.
        However, this approach may not generalise to binary systems where the primary has a significant signature in the light curves.
        Under such a scenario, tweaks to both primary rotation and secondary orbit periods may need to be made that permit a spin-orbit resonance but only modify the light curve signatures at a level less than the scatter in the IR observations.
        If such modifications are not possible without significantly affecting the match to thermal-IR light curves, then an alternative modelling approach would need to be found (see Section~\ref{sec:future} for additional discussion).

        For each target, an initial estimate of the Bond albedo was required as an input to the thermophysical model.
        To derive this, phase curve data were used to fit the $HG$ photometric system \citep{1989aste.conf..524B} to get the absolute magnitude, $H$, and the slope parameter, $G$.
        For (3905) Doppler, the $HG$ parameters were derived from a fit to V-band data from the Minor Planet Center\footnote{\url{https://www.minorplanetcenter.net}} due to a lack of available high-quality photometric data.
        The fits to these data result in parameters derived as $H = 12.657 \pm 0.026$ and $G = 0.215 \pm 0.025$.
        For (175706) 1996 FG3, the $HG$ parameters were taken as $H = 17.833 \pm 0.024$, $G = -0.041 \pm 0.005$ from \citet{2011MNRAS.418.1246W} as these were based on better quality data than those reported by the Minor Planet Center.
        
        At this stage, the \textit{WISE} data can be fit using a simple thermal model such as the NEATM \citep{1998Icar..131..291H} to derive an estimate of the effective diameter of the binary systems.
        The geometric albedo, $p_\text{V}$ can then be calculated from the fitted diameter and the absolute magnitude as
        \begin{equation}
            p_\text{V} = \left(\frac{1329 \times 10^{-H/5}}{D_\text{eff.}}\right)^2
        \end{equation}
        \citep{irasChap4}.
        From this the effective Bond albedo can be calculated as
        \begin{equation}
            A_\text{B\_EFF} = q p_\text{V} = (0.290 + 0.684 G) p_\text{V}.
        \end{equation}
        This is not the Bond albedo we can use in the model, however, as this is the effective rough-surface Bond albedo and the thermophysical model requires the smooth-surface Bond albedo as input.
        For the case of the hemispherical crater representation of surface roughness, the smooth surface Bond albedo can be derived from the effective Bond albedo and the roughness fraction by inversion of Equation~\ref{eq:bondAlbedo}
        \begin{equation}\label{eq:bondAlbedo}
            A_\text{B\_EFF} = f_R\frac{A_\text{B}}{2-A_\text{B}}+(1-f_R)A_\text{B}
        \end{equation}
        \citep{2011MNRAS.418.1246W}.
        This conversion requires assumptions about the roughness fraction, and also assumes a good fit using the NEATM for the diameter.
        Therefore the input Bond albedo is calculated using a roughness fraction of $f_R = 0.5$, and a correction factor is introduced during fitting to `correct' for the differing flux predictions by models with varying diameter and roughness fraction.
        This approach allows for a single value of the Bond albedo to be input into the model, instead of leaving it as a free parameter (which would result in excessive computation time).
        Provided that this flux correction is within 10 per cent of unity, then the input model Bond albedo is assumed to provide a reasonable approximation.
        The Bond albedos input into the model are: $0.137$ for (3905) Doppler (derived from $H$, $G$ parameters), and $0.01$ for (175706) 1996 FG3 \citep[value from][]{2011MNRAS.418.1246W}.

\section{Eclipse and Occultation Events in Disk-Integrated Data}\label{sec:fitting}

    \subsection{(3905) Doppler}

        The WISE Cryogenic observations of (3905) Doppler from 2010 March 06 contained data in all of the WISE channels.
        As the data in the shorter wavelength W1 and W2 channels contain a significant amount of reflected flux, we only use the data from the W3 and W4 channels (which have negligible reflected flux contribution) in the modelling process for this object.
        In the thermophysical model fits to the (3905) Doppler WISE data there were 4 free parameters: diameter, thermal inertia, roughness fraction, and rotational phase.
        The additional rotational phase parameter was required as the rotational state of the system at the time of the observations was unknown.
        The IR light curves showed clear rotational/orbital variability, allowing for the phase to be constrained instead of using rotationally averaged fluxes in the fitting process.

        The thermophysical model fit to the data results in parameter estimates as follows: $D_\text{eff.} = 7.971 \pm 0.119\,\si{\kilo\metre}$, $\Gamma = 114 \pm 31\,\tiu$, $f_R = 0.73 \pm 0.22$, with a reduced chi-squared value of $\chi^2_\nu = 0.94$.
        The fitted diameter is an effective diameter of the two bodies combined, which can be converted to area-equivalent component diameters of $D_1 = 5.82 \pm 0.08\,$km and $D_2 = 5.47 \pm 0.08\,$km assuming that $D_1^2 + D_2^2 = D_\text{eff.}^2$ \citep{2012ApJ...748..104W}.
        Assuming that the main-belt binaries observed and modelled using a TPM in \citet{2012Icar..221.1130M} are dominated by particulate regolith rather than porous boulders, we can compare their thermal inertia values to that of (3905) Doppler by correcting for the differences in thermal inertia due to heliocentric distance \citep{2018MNRAS.477.1782R}.
        Making this correction to the \citet{2012Icar..221.1130M} values to derive the thermal inertia ranges as they would be at the modelled heliocentric distance of (3905) Doppler, the thermal inertia derived as part of this work is consistent within uncertainty with values derived for 5 of the 7 binaries in \citet{2012Icar..221.1130M}.
        
\begin{figure*}
            \centering
            \begin{subfigure}{0\textwidth}
                \phantomsubcaption
                \label{fig:dopplerDataFita}
            \end{subfigure}
            \begin{subfigure}{0\textwidth}
                \phantomsubcaption
                \label{fig:dopplerDataFitb}
            \end{subfigure}
            \begin{subfigure}{0\textwidth}
                \phantomsubcaption
                \label{fig:dopplerDataFitc}
            \end{subfigure}
            \includegraphics[width=\textwidth]{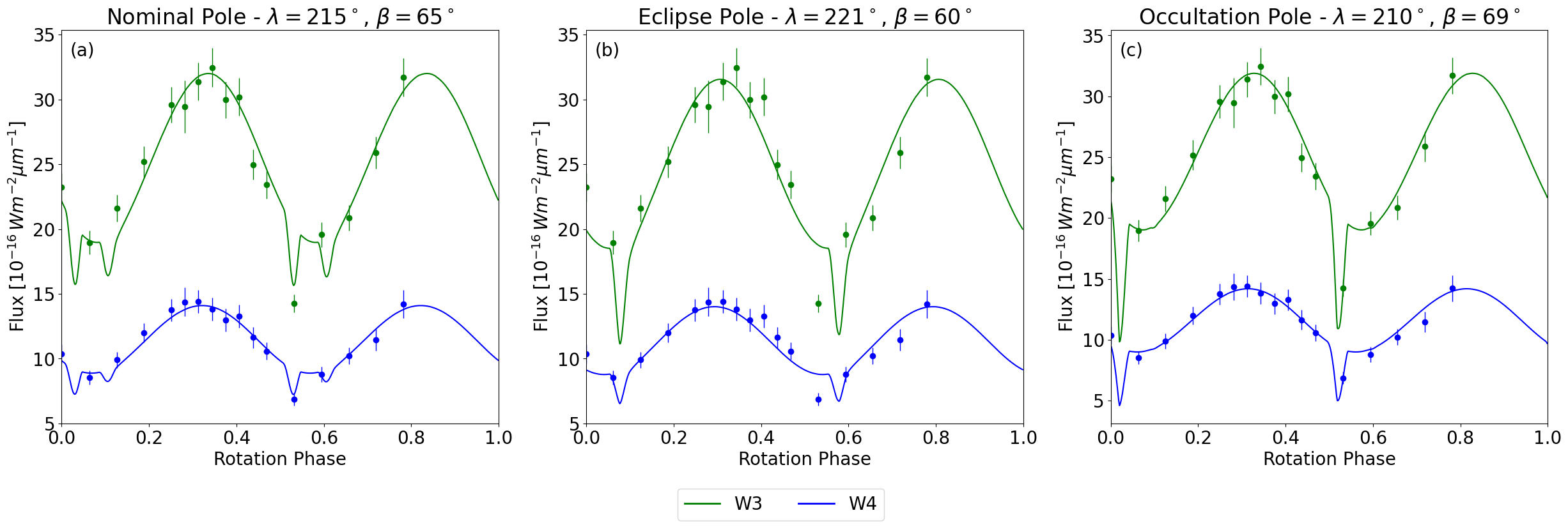}
            \caption{Cryogenic \textit{WISE} data for (3905) Doppler, overplotted with the ATPM model fits. \textbf{(a)} Fit for the nominal mutual orbit pole. \textbf{(b)} Fit for the pole that maximises eclipses and minimises occultations. \textbf{(c)} Fit for the pole that minimises eclipses and maximises occultations.}
            \label{fig:dopplerDataFit}
        \end{figure*}

\begin{figure*}
            \centering
            \begin{subfigure}{0\textwidth}
                \phantomsubcaption
                \label{fig:dopplerChi2a}
            \end{subfigure}
            \begin{subfigure}{0\textwidth}
                \phantomsubcaption
                \label{fig:dopplerChi2b}
            \end{subfigure}
            \begin{subfigure}{0\textwidth}
                \phantomsubcaption
                \label{fig:dopplerChi2c}
            \end{subfigure}
            \includegraphics[width=\textwidth]{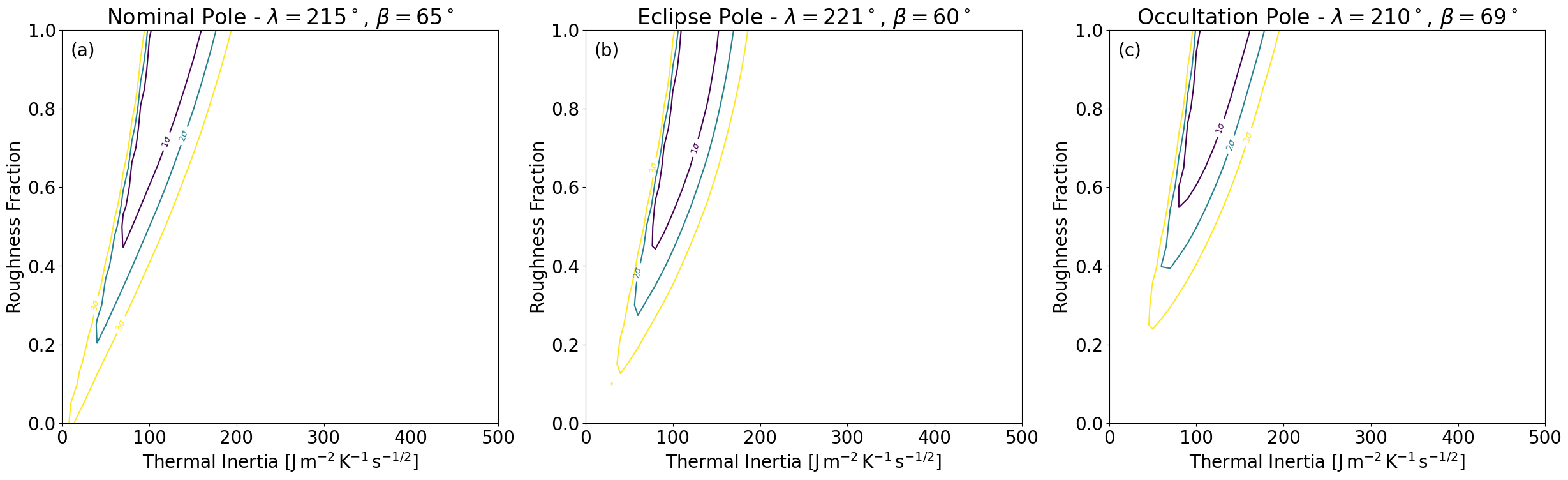}
            \caption{Roughness versus thermal inertia chi-squared planes for ATPM fits to the (3905) Doppler cryogenic \textit{WISE} data. \textbf{(a)} Fit for the nominal mutual orbit pole. \textbf{(b)} Fit for the pole that maximises eclipses and minimises occultations. \textbf{(c)} Fit for the pole that minimises eclipses and maximises occultations.}
            \label{fig:dopplerChi2}
        \end{figure*}

        The thermal inertia value of $\Gamma = 114 \pm 31\,\tiu$ suggests that the surface of (3905) Doppler is dominated by the presence of fine-grained regolith, although low thermal inertia values on the C-type asteroids Bennu and Ryugu can also be explained by the presence of highly porous boulders \citep{2020SciA....6.3699R}.
        It is not yet known if S-type asteroids such as (3905) Doppler can have surfaces comprising porous boulders, and so the fine-regolith scenario is considered the most plausible at this stage.
        The model fits to the W3 and W4 data using the `nominal pole' solution from \citet{2020Icar..34513726D} can be found in Figure~\ref{fig:dopplerDataFita}.
        A $\chi^2$ contour plane for thermal inertia and roughness fraction can be found in Figure~\ref{fig:dopplerChi2a} for the nominal pole solution.
        
        The thermal-IR light curves produced by the model predict that there should be both eclipses and occultations in the data.
        There is insufficient data to detect any of these events at the first set of expected mutual events.
        The second set of mutual events in the simulation has data that should lie in both events, although we only see a data point in the dip from the first mutual event of the pair (and the depth of the predicted event is too low).
        From this we assessed that it was possible that the pole solution could be slightly different, and tried two different pole solutions: one that removed occultations and maximised eclipse depth (the `Eclipse Pole'), and one that removed eclipses and maximised occultation depth (the `Occultation Pole').
        The data were then re-fit using these pole solutions and the results of this re-analysis can be found in Figures~\ref{fig:dopplerDataFitb} and \ref{fig:dopplerChi2b} for the `Eclipse Pole' and in Figures~\ref{fig:dopplerDataFitc} and \ref{fig:dopplerChi2c} for the `Occultation Pole'.
        The fit to the data using the `eclipse pole' gave parameters as follows: $D_\text{eff.} = 7.953 \pm 0.101\,\si{\kilo\metre}$, $\Gamma = 121 \pm 26\,\tiu$, $f_R = 0.74 \pm 0.20$, with a reduced chi-squared value of $\chi^2_\nu = 3.82$.
        The fit to the data using the `occultation pole' gave parameters as follows: $D_\text{eff.} = 7.978 \pm 0.117\,\si{\kilo\metre}$, $\Gamma = 122 \pm 26\,\tiu$, $f_R = 0.79 \pm 0.17$, with a reduced chi-squared value of $\chi^2_\nu = 0.62$.
        
        We see from these re-analyses that a pole that maximises eclipse and removes occultations is not well supported by the data as the event timing is not correct, although an `occultation pole' is supported by the data as the data aligns better with the model.
        Due to the $\chi^2_\nu$ value of the `occultation pole' fit being further from unity than the `nominal pole' fit, we can not rule out that either the model has been over-fit or that the uncertainties are over-estimated.
        Without denser light curve coverage it is not possible to constrain whether the `occultation pole' is actually better than the `nominal pole' as only one mutual event is detected in the data over the whole rotational phase of the system.

        To assess the sensitivity of our results to the pole orientation, we constructed a grid of nine pole solutions across the 3-sigma range of the nominal pole solution (i.e. pole latitudes of 59, 65, and 71 degrees, and pole longitudes of 209, 215, and 221 degrees).
        The model was then run for each pole in the grid to find the best fit thermal inertia under those geometries.
        The standard deviation of these fitted values is \textasciitilde$7\,\tiu$, which only increases the total uncertainty by approximately 3 per cent when added in quadrature to the uncertainties in the original fit.
        This small effect indicates that with such low uncertainties in the pole solution (on the order of a few degrees), a good fit can be obtained using a model like the one presented in this work.
        However, care should be taken when analysing objects where the uncertainties are more significant (or where degenerate solutions exist from modelling processes).
        
        The different types of thermal-IR light curves that can be obtained for fully synchronous systems is apparent, however, in these figures.
        An animation demonstrating the geometry of the system for each trial pole along with their light curves can be found at \url{https://doi.org/10.6084/m9.figshare.25486924}.
        
        It can be seen across all trial poles that the roughness fraction is poorly constrained.
        This is due to the single geometry at which we have modelled this target; for a good constraint of roughness a number of different phase angles should be sampled to constrain the presence of IR beaming which is enhanced through the presence of surface roughness.
        Additional cryogenic \textit{WISE} data were not available for this object, and since this work is not focused on determining the properties of (3905) Doppler and instead focuses on the model and the effect of mutual events in thermal-IR data, we do not seek to model this asteroid using non-cryogenic NEOWISE W2 data which would require estimation of the reflected flux before thermophysical modelling.


    \subsection{(175706) 1996 FG3}\label{subsec:FG3}

        To model (175706) 1996 FG3, we obtain data from both the WISE Cryogenic period and also the post-cryo NEOWISE period of operations.
        As this asteroid is a C-type asteroid, the contribution of reflected flux in the W2 band is negligible compared to the uncertainties in the data, and so the W2 fluxes can be modelled as purely thermally emitted flux.
        NEOWISE data were obtained that were taken at heliocentric distances within $0.1\,$au of the heliocentric distance of (175706) 1996 FG3 during the cryogenic WISE observations, in order to avoid the effect of variable thermal inertia with heliocentric distance \citep{2018MNRAS.477.1782R}.
        In the thermophysical model fits to the (175706) 1996 FG3 data there were 3 free parameters: diameter, thermal inertia, and roughness fraction.
        Unlike (3905) Doppler, the thermal-IR data quality was insufficient to constrain the rotational phase as a free parameter.
        Therefore we used the shape model of (175706) 1996 FG3 and a Lambert/Lommel-Seeliger scattering model \citep[as implemented by][]{2001Icar..153...24K} to simulate its light curve at the same positions as data reported by \citet{2015Icar..245...56S}.
        The rotation phase offset was then permitted to vary, along with minor adjustments to the rotation period, the find the optimum combination that matches the \citet{2015Icar..245...56S} photometric data.
        The results of this offset vs. period scan can be found in Figure~\ref{fig:FG3phasing}, with the best solution found with a phase offset of 229 degrees and a slight modification of the rotation period to $P = 16.15076$ hours.
        The scattering model fits to the photometric data are shown in Figure~\ref{fig:optical_fits}, showing good agreement between the model and the data.

\begin{figure*}
            \centering
            \includegraphics[width=.825\textwidth]{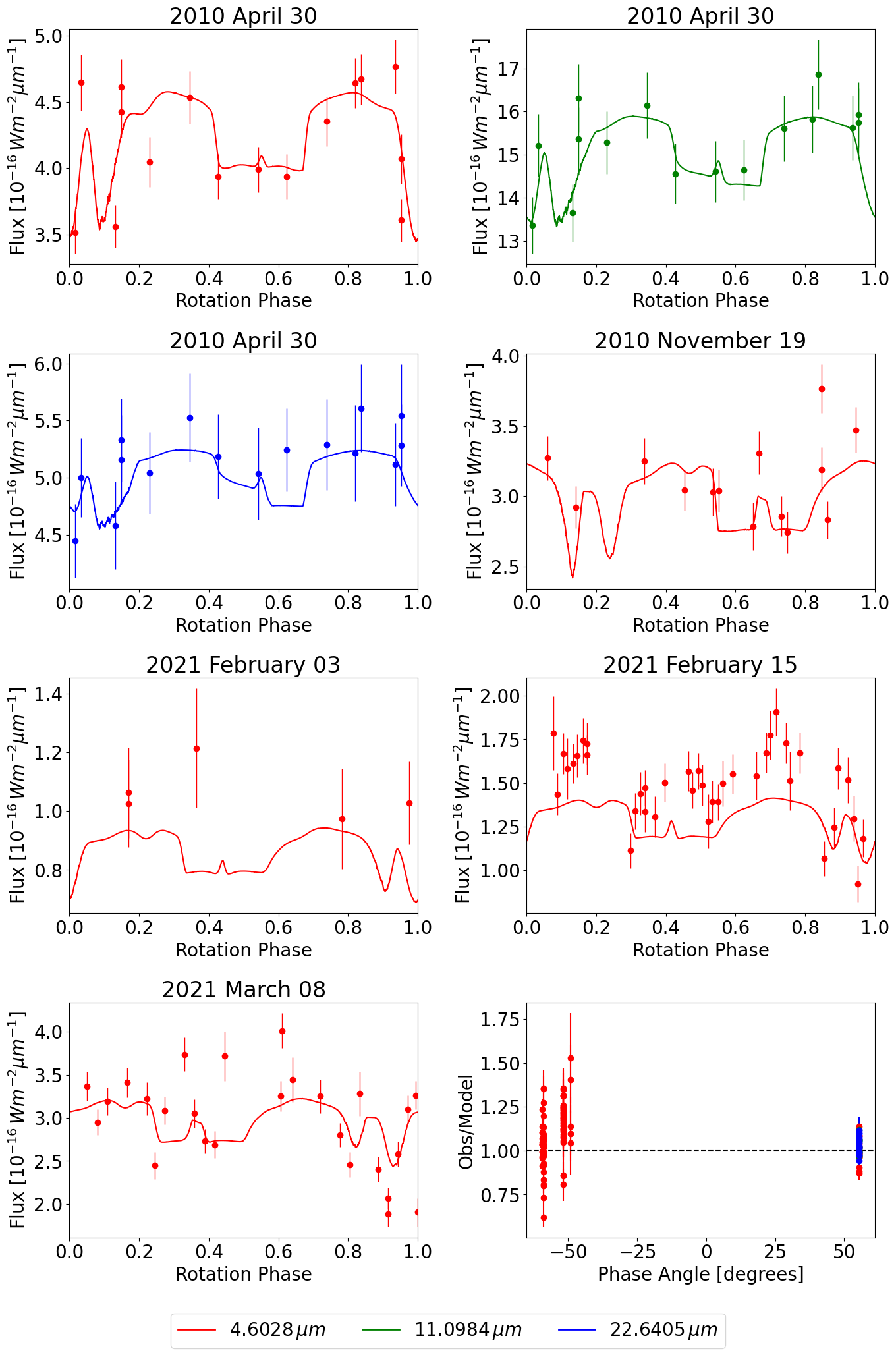}
            \caption{\textit{WISE} data of (175706) 1996 FG3 used in the thermophysical modelling in this work, alongside the fitted model predicted light curves in each wavelength (where appropriate). A systematic offset in the February 2021 data is unresolved and can not be explained by the presence of reflected flux in the W2 data.}
            \label{fig:FG3fits}
        \end{figure*}

        The thermophysical model fit to the \textit{WISE} and \textit{NEOWISE} data results in parameter estimates as follows: $D_\text{eff.} = 1.554 \pm 0.019\,$km, $\Gamma = 142 \pm 6\,\tiu$, $f_R = 0.31 \pm 0.06$, with a reduced chi-squared value of $\chi^2_\nu = 3.65$. 
        This $\chi^2_\nu$ value is high, indicating that there is a systematic issue with either the data or the model fit.
        We suggest the high value of the $\chi^2_\nu$ is caused by an under-estimation of the uncertainty in these W2 data, as the uncertainties are on average 8 per cent but the scatter in the data around the model is around 13 per cent.
        Upon visual inspection of the raw WISE images\footnote{Available at: \url{https://irsa.ipac.caltech.edu/applications/wise/}}, it can be seen that (175706) 1996 FG3 passes through some crowded star fields during 2021, which may have affected the quality of these latter datasets. 
        It can be seen in some of the model fits to the W2 data (Figure~\ref{fig:FG3fits}), there appear to be some systematic offsets between the data and the model.
        The presence of reflected flux is evaluated to not be the cause of the offset, as the data presented in Figure~\ref{fig:FG3fits} have had the reflected flux contribution estimated and removed using the methods of \citet{2018MNRAS.477.1782R}, which provided a negligible shift in the data.
        The cause of this systematic offset is undetermined in this work, and the results reported for (175706) 1996 FG3 should be interpreted with care.
        In the final panel of Figure~\ref{fig:FG3fits} we present the ratio between the observed \textit{WISE} data and the ATPM model fit as a function of phase angle (where negative phase angle corresponds to seeing the morning side of the asteroid in this case).
        It is seen that there is a large degree of scatter around the model for observations of the morning side.
        Not all of the observations at negative phase angles are systematically brighter than the model, so errors in modelling the morning/afternoon temperature distribution can not explain the systematic offsets we see in the February 2021 data.
        The W2 channel is less sensitive to the cooler temperatures expected on the morning side of the asteroid, and so this may be limiting the signal-to-noise in the observations and causing the increase in scatter around the model for observations at negative phase angles.
        There also exists the potential that local topography on the surface of the primary could influence the fits to the light curves.
        In this work we have assumed a shape for the primary that may not capture subtle light curve effects that could be produced by craters or other non-convex features.
        However, the limiting factor in the analysis of (175706) 1996 FG3 is the low signal-to-noise in the observations.
        Application of this thermal model to flyby or orbiter data such as that expected from the Hera mission (see Section~\ref{sec:hera} for further discussion) will require excellent local topography models to ensure systematic effects from assumed shapes do not dominate the uncertainty in the model fits.
        
\begin{figure}
            \centering
            \includegraphics[width=\linewidth]{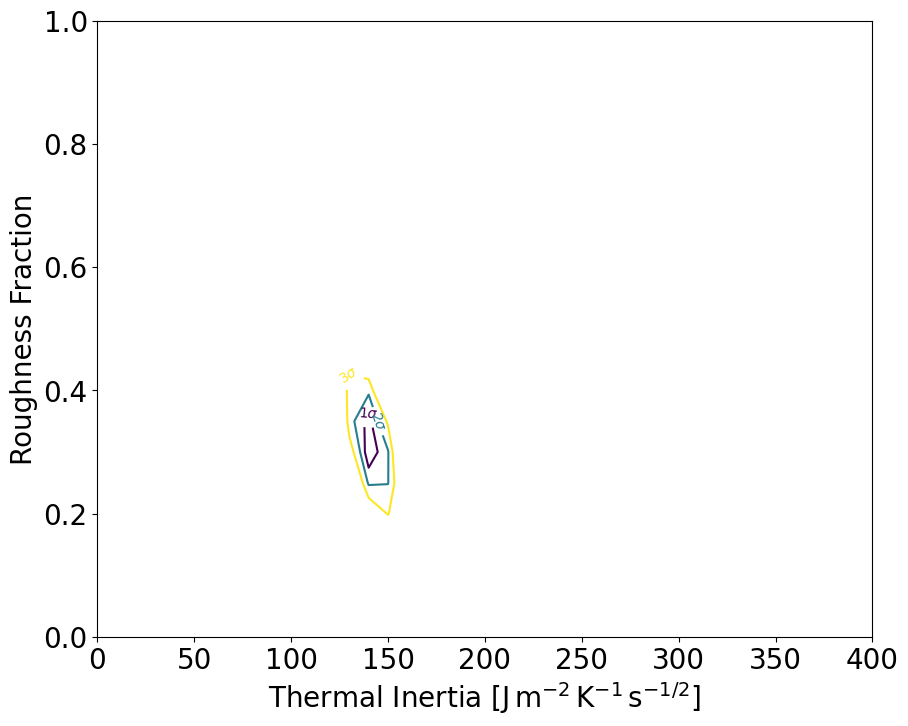}
            \caption{Roughness versus thermal inertia chi-squared contours showing the well-constrained fit of these parameters using the \textit{WISE} data of (175706) 1996 FG3.}
            \label{fig:FG3results}
        \end{figure}

        The fitted diameter is an effective diameter of the two bodies combined, which can be converted to area-equivalent component diameters of $D_1 = 1.49 \pm 0.02\,$km and $D_2 = 0.43 \pm 0.01\,$km assuming that $D_1^2 + D_2^2 = D_\text{eff.}^2$ \citep{2012ApJ...748..104W}.
        An animation demonstrating the geometry of the system at the cryogenic \textit{WISE} observation time along with the light curves in the W2, W3, and W4 bands can be found at \url{https://doi.org/10.6084/m9.figshare.25486924}.
        A $\chi^2$ contour plane for thermal inertia and roughness fraction can be found in Figure~\ref{fig:FG3results}.
        
        The thermal inertia derived in this work is comparable within the uncertainty with a value of $\Gamma = 120 \pm 50\,\tiu$ derived by \citet{2011MNRAS.418.1246W}.
        This value for the thermal inertia is comparable to that of (3905) Doppler once heliocentric distance is accounted for, and therefore is also consistent with the majority of the thermal inertia values derived for main-belt binaries by \citet{2012Icar..221.1130M}.
        As with (3905) Doppler, the low thermal inertia derived in this work suggests that (175706) 1996 FG3 has a surface dominated by either fine-grained regolith or porous boulders.
        The thermal inertia would be expected to remain constant for porous boulders \citep{2020SciA....6.3699R}, and vary with heliocentric distance for fine-grained regolith \citep{2018MNRAS.477.1782R}.
        This degeneracy is unable to be broken with the current corpus of data, due to a lack of measurements at significantly different heliocentric distances that would permit evaluating the temperature dependence of the thermal inertia.

\section{Sensitivity of Binary Thermal-IR Light Curves to Thermophysical Parameters}\label{sec:sensitivity}

    With the binary thermophysical model demonstrated to be able to reproduce and account for mutual events in the thermophysical modelling process, we now focus on how different parameters affect binary thermal-IR light curves.
    To achieve this, we will focus on producing simulated thermal-IR light curves of (175706) 1996 FG3 at the viewing geometries of the \citet{2015Icar..245...56S} data.
    The parameters varied in this study were: thermal inertia, `observed' wavelength, and roughness fraction.
    
\begin{figure}
        \begin{subfigure}{\linewidth}
            \centering
            \includegraphics[width=.95\linewidth]{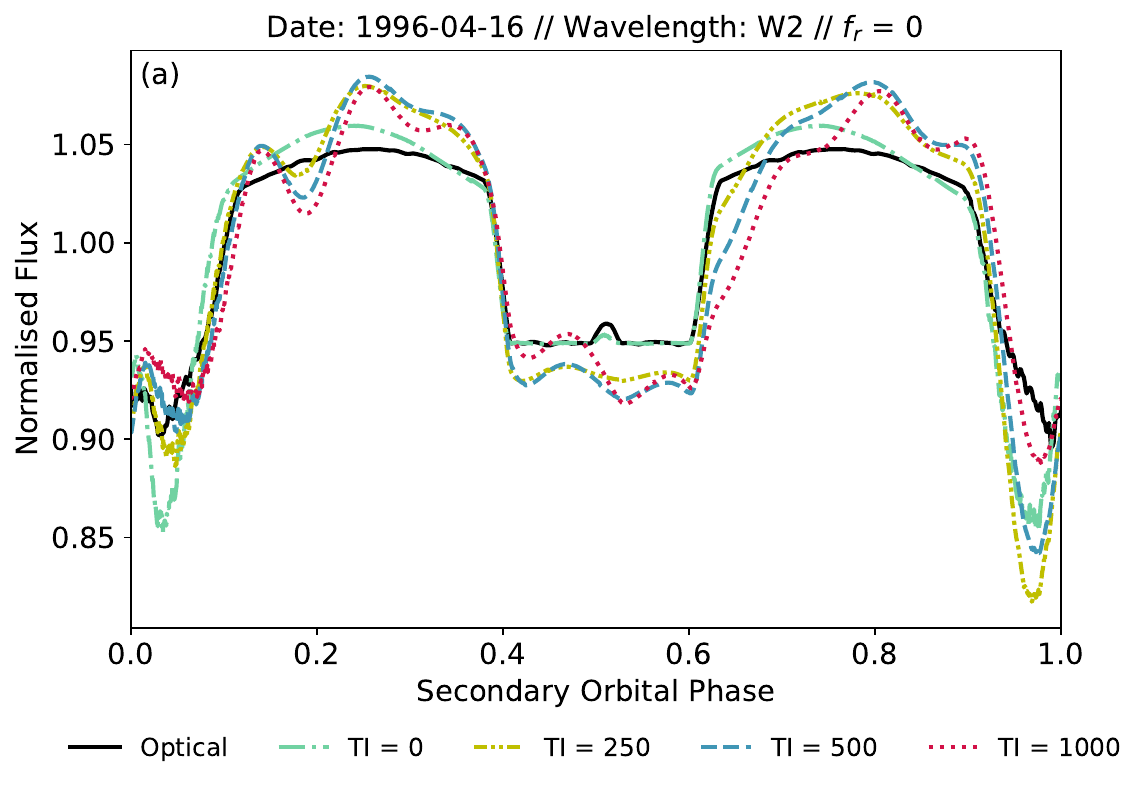}
            \phantomsubcaption
            \label{fig:TI_example}
        \end{subfigure}\\
        \begin{subfigure}{\linewidth}
            \centering
            \includegraphics[width=.95\linewidth]{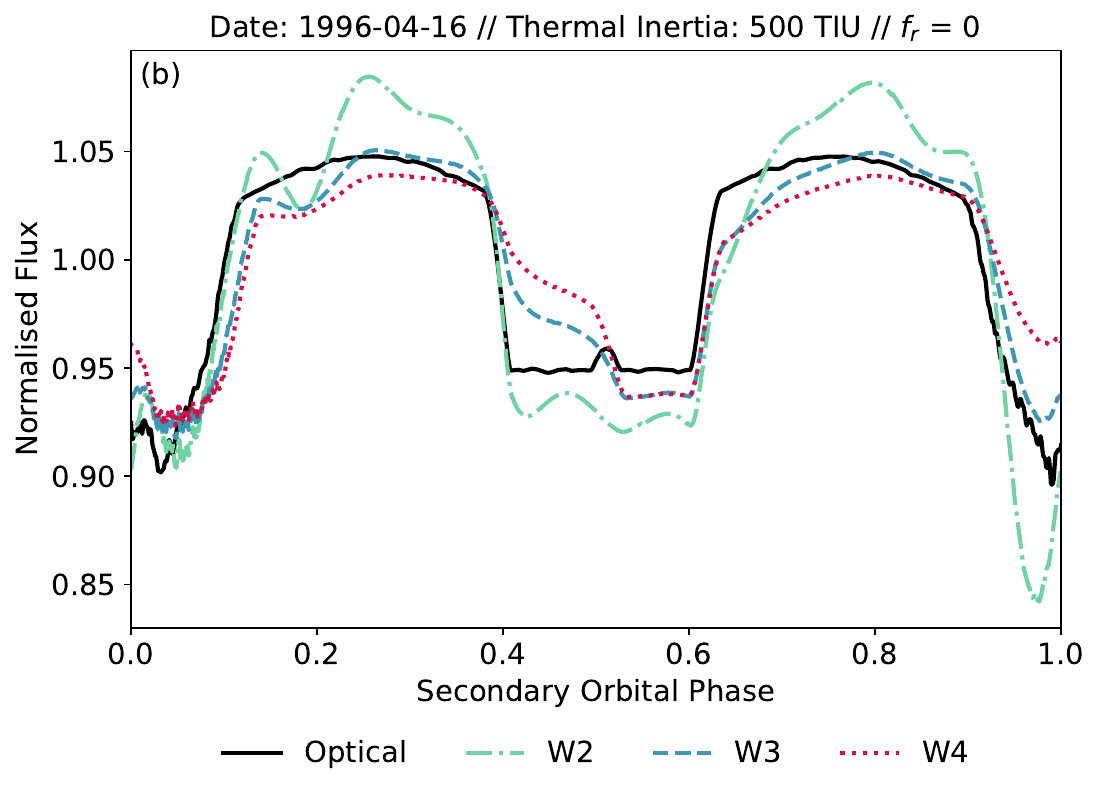}
            \phantomsubcaption
            \label{fig:Wavelength_example}
        \end{subfigure}\\
        \begin{subfigure}{\linewidth}
            \centering
            \includegraphics[width=.95\linewidth]{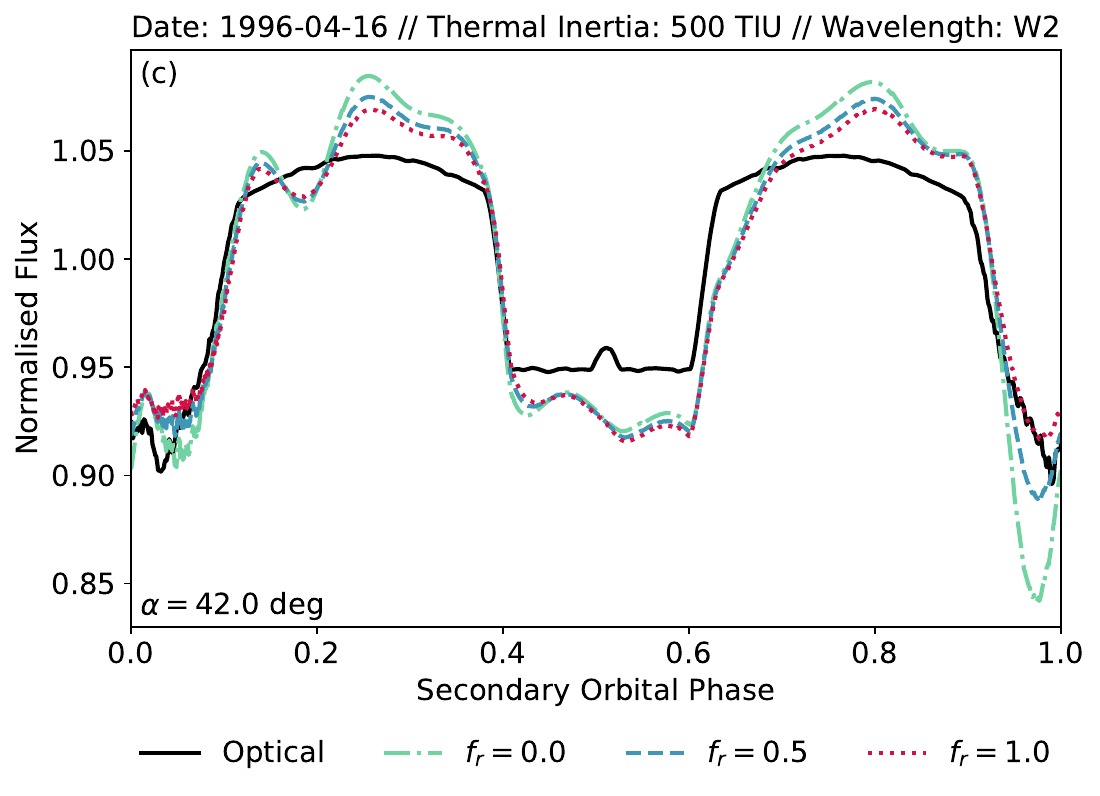}
            \phantomsubcaption
            \label{fig:Roughness_example}
        \end{subfigure}
        \caption{Example plots from Figures~\ref{fig:ti_study}-\ref{fig:roughness_study} showing how the morphology of the thermal-IR light curve depends on: \textbf{(a)} thermal inertia, \textbf{(b)} wavelength of observation, and \textbf{(c)} thermal roughness.}
        \label{fig:parameterStudyExamples}
    \end{figure}

    The effect of changing thermal inertia on the light curves is shown across all of the simulated geometries in Figure~\ref{fig:ti_study}, with an example shown for one geometry in Figure~\ref{fig:TI_example}.
    In these simulations we trial four values for the thermal inertia: $0$, $250$, $500$, and $1000\,\tiu$.
    A simulated optical reflectance light curve is included for comparison in these figures.
    In these simulations the roughness fraction is held fixed at 0 (i.e. a completely smooth surface), with the light curve `observed' in the \textit{WISE} W2 ($4.6\,\si{\micro\metre}$) band.
    
\begin{figure}
        \centering
        \includegraphics[width=\linewidth]{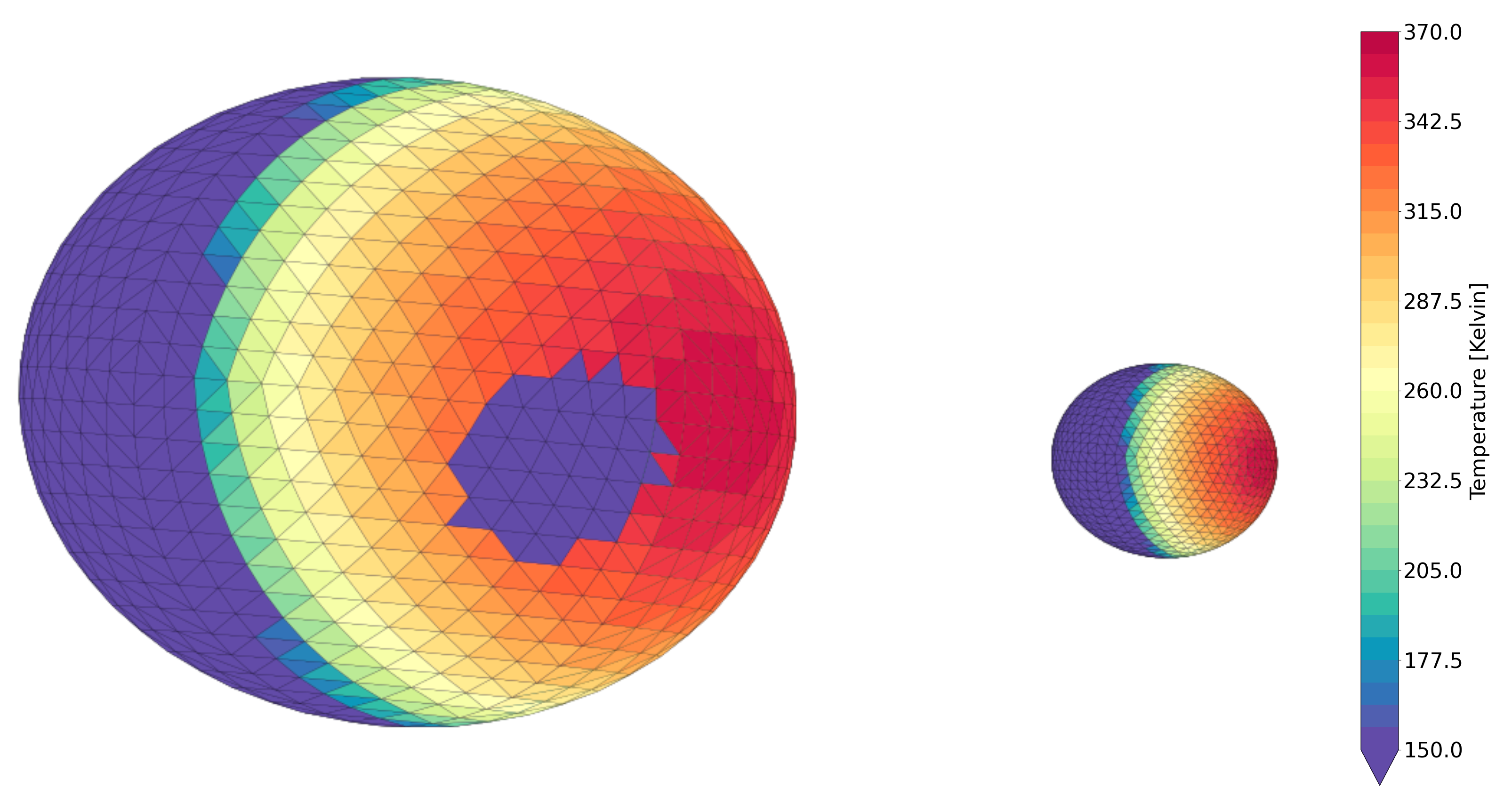}
        \includegraphics[width=\linewidth]{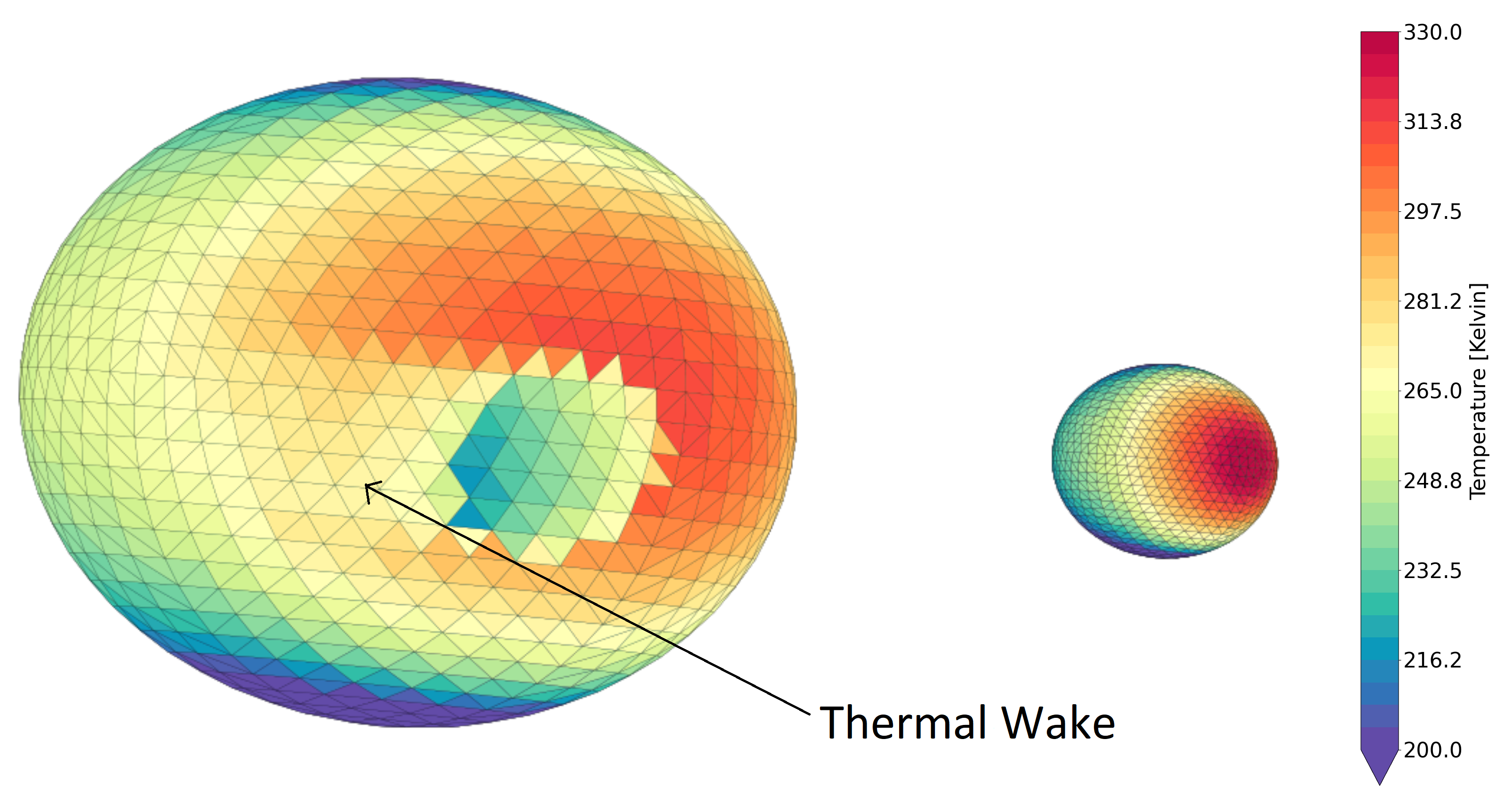}
        \caption{Snapshots of the surface temperature distribution of (175706) 1996 FG3 simulated for the 2010 April 30 \textit{WISE} viewing geometry, demonstrating how at non-zero thermal inertia values the faster rotation of the primary compared to the secondary orbit causes a thermal wake to be formed ahead of the eclipse shadow. \textbf{Top Panel:} Simulated with a thermal inertia of $0\,\tiu$, no thermal wake is seen as the surface is in instantaneous equilibrium. \textbf{Bottom Panel:} Simulated with a thermal inertia of $400\,\tiu$, a thermal wake forms as surface elements rotated out of the eclipse shadow take time to re-heat to their non-shadowed equivalent temperature.}
        \label{fig:FG3_wake_comps}
    \end{figure}

    In these light curves, it is clear that increasing thermal inertia introduces significant modulations to the light curves compared to the zero thermal inertia case.
    This modulation arises from the non-synchronous orbit of the secondary of (175706) 1996 FG3.
    The primary body of FG3 rotates faster than the secondary orbits, and so any elements of the surface that were shadowed at one time-step can then become unshadowed at the next time-step.
    With non-zero thermal inertia, these surface elements remain cooler than nearby elements that were not shadowed by the secondary, and take time to heat up to the temperature they would otherwise be if they had not been shadowed (referred to hereafter as the `unshadowed equivalent temperature').
    As new surface elements are shadowed by the secondary, these also experience the same cooling and slow heating, creating a stretched cold spot on the surface of the primary (which we refer to as the `thermal wake' hereafter).
    The lack of a thermal wake for instantaneous equilibrium can be seen in the top panel of Figure~\ref{fig:FG3_wake_comps}, and the presence of this effect can be seen for a thermal inertia value of $400\,\tiu$ in the bottom panel.
    This thermal wake returns to the observer line-of-sight after one primary rotation (but without being shadowed again due to the progression of the orbit of the secondary).
    If the thermal wake has not completely heated back up to the unshadowed equivalent temperature, then this changes the disk-averaged temperature of the surface visible by the observer, and modulates the flux observed.
    For large values of the thermal inertia it is seen that this thermal wake can have a significant impact on the morphology of a thermal-IR light curve, and it is therefore important to include these effects when applying thermophysical models to high-quality data of binary asteroid systems.

\begin{figure*}
        \centering
        \includegraphics[width=\textwidth]{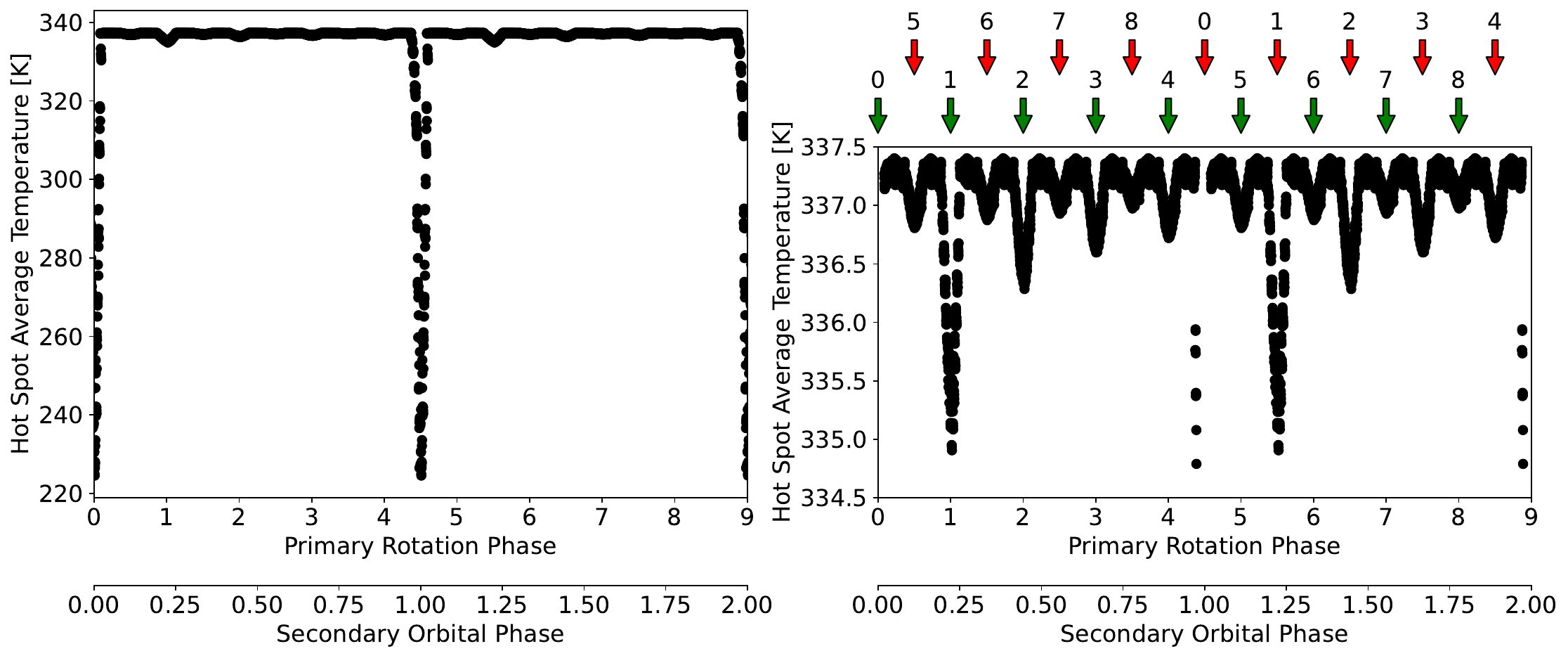}
        \caption{ \textbf{Left Panel:} Plot showing the evolution of the temperature of the hot spot on the surface of (175706) 1996 FG3 over two orbits of the secondary. \textbf{Right Panel:} Same as the left panel but with the y-axis range modified to see the smaller-scale variations in temperature. Green arrows are pointing to times when the first thermal wake passes over the hot spot, and the red arrows point to the times when the second thermal wake passes over the hot spot. the second thermal wake occurs due to a second eclipse event 4.5 primary rotations after the first eclipse event.}
        \label{fig:hotspot}
    \end{figure*}

    Depending on the thermal inertia, this thermal wake can then still be detected after multiple rotations of the primary until the surface experiences another eclipse event by the secondary.
    Using this example of (175706) 1996 FG3, which has the primary rotating 9 times for every 2 orbits of the secondary, we can see 2 thermal wakes on the surface as a second eclipse shadowing occurs 4.5 primary rotations after the first.
    This can be observed in Figure~\ref{fig:hotspot}, which tracks the temperature of the `hot spot' on the surface of the primary of (175706) 1996 FG3 over 9 primary rotation periods (2 secondary orbital periods).
    The effect of the thermal wake on the temperature of the hot spot is small for later primary rotations, and would likely be hidden by the noise from \textit{WISE} or other telescope data.
    However, this effect may be significant during high-resolution imaging in the thermal-IR by spacecraft visiting binary asteroid systems.
    The applicability of our binary TPM to missions such as ESA's Hera mission and NASA's Lucy mission, where the effect of the thermal wake may be significant in the data, is discussed further in Section~\ref{sec:discussion}.
    
    The effect of changing the wavelength of observation on the light curves is shown across all of the simulated geometries in Figure~\ref{fig:wavelength_study}, with an example shown for one geometry in Figure~\ref{fig:Wavelength_example}.
    In these simulations we use three wavelengths of observation, corresponding to the three longest wavelength bands of \textit{WISE}: W2 ($4.6\,\si{\micro\metre}$), W3 ($12\,\si{\micro\metre}$), W4 ($22\,\si{\micro\metre}$).
    A simulated optical reflectance light curve is included for comparison in these figures.
    In these simulations the roughness fraction is held fixed at 0 (i.e. a completely smooth surface), with a thermal inertia of $500\,\tiu$.
    The predominant features seen when varying the wavelength of observation during these simulations is that the effect of the thermal wake is most significant in the W2 `observations', and is less pronounced at longer wavelengths.
    The likely explanation for this is that the shadowing and subsequent thermal wake at these geometries resulted in more significant temperature drops on the hot-spot of the surface (i.e. near the area of maximum insolation).
    With the hotter part of the surface affected by these repeated cooling events, the shorter wavelength observations that are more sensitive to the hotter parts of the surface will be modulated more by the thermal wake.
    It is also seen that the morphology of eclipse events is dependent on the wavelength of observation, largely resulting from the sensitivity of the longer wavelength channels to cooler temperatures (and so the large drop in temperature of the hot spot from an eclipse event is not as significant in these channels compared to in the W2 channel).
    Occultation event morphology, however, is seemingly not affected by wavelength of observation.
    The differences between some of the occultation depths in the simulated data likely arise from the variable affect of the thermal wake with wavelength as discussed previously. 
    
    The effect of changing roughness on the light curves is shown across all of the simulated geometries in Figure~\ref{fig:roughness_study}, with an example shown for one geometry in Figure~\ref{fig:Roughness_example}.
    In these simulations we try three roughness fractions of $0.0$, $0.5$, and $1.0$, corresponding to RMS slope values of $0\si\degree$, $35\si\degree$, and $49\si\degree$ respectively.
    A simulated optical reflectance light curve is included for comparison in these figures.
    In these simulations the thermal inertia is held fixed at $500\,\tiu$, with the light curve `observed' in the \textit{WISE} W2 ($4.6\,\si{\micro\metre}$) band.
    The effect of roughness in these simulated thermal-IR observations is hard to distinguish.
    Across the majority of the light curves there are minimal differences out of mutual events, although there is some evidence that mutual event depths may have some level of dependence on the roughness.
    It is unclear, however, if this is an artefact of placing roughness elements onto the low-resolution shape model we adopted for (175706) 1996 FG3 and the subsequent change to the viewing geometry calculations for each facet.
    The differences are seemingly most pronounced at low phase angles, suggesting a potential physical effect similar to IR beaming at low phase angles having an effect on mutual event depth.

\section{Discussion}\label{sec:discussion}

    \subsection{Application of the Binary TPM to ESA's Hera Mission}\label{sec:hera}

        As discussed previously, the binary TPM presented in this work can be applied to synchronous and semi-synchronous binary systems in disk-integrated IR telescope observations.
        Another potential application is for analysis of thermal-IR images and spectra taken from upcoming missions to binary asteroids (if the targets have synchronous or approximately semi-synchronous secondary orbits).
        ESA's upcoming Hera mission \citep{2022PSJ.....3..160M} to the binary asteroid (65803) Didymos is the follow-up to NASA's Double Asteroid Redirect Test (DART) mission \citep{2021PSJ.....2..173R,2024PSJ.....5...49C} which provided the first test of a kinetic impact deflection of an asteroid.
        The Hera mission will carry a thermal-infrarer imager (TIRI) which will have an absolute temperature accuracy of \textasciitilde$ 3\,\si\kelvin$\footnote{\url{https://www.heramission.space/hera-instruments}}.
        The accuracy of TIRI means that `thermal wakes' present on the surface of Didymos could be seen near to the limit of the instrument.
        Due to the proximity of the Didymos rotation period and Dimorphos orbital period to a 5:1 relationship \citep[$\approx 5.03:1$ post-impact;][]{2023Natur.616..448T}, it may be possible to apply the model presented in this work to interpret data from the TIRI instrument on the Hera mission.

        The ATPM has previously been used to estimate the thermal inertia and roughness of Didymos and Dimorphos \citep{2023PSJ.....4..214R,2024PSJ.....5...66R}, although those analyses did not include the effect of mutual events in the analysis.
        Due to the high resolution shape models currently available, which will be further refined as part of the Hera mission, we provide some approximations that could be made to model this system during Hera operations with minimal computational expense.
        As Hera will be able to image each component of the binary separately, rather than their fluxes being combined in disk-integrated data, we can model the objects individually while including the effects of the other component on the target of interest.

\begin{figure}
            \centering
            \includegraphics[width=\linewidth]{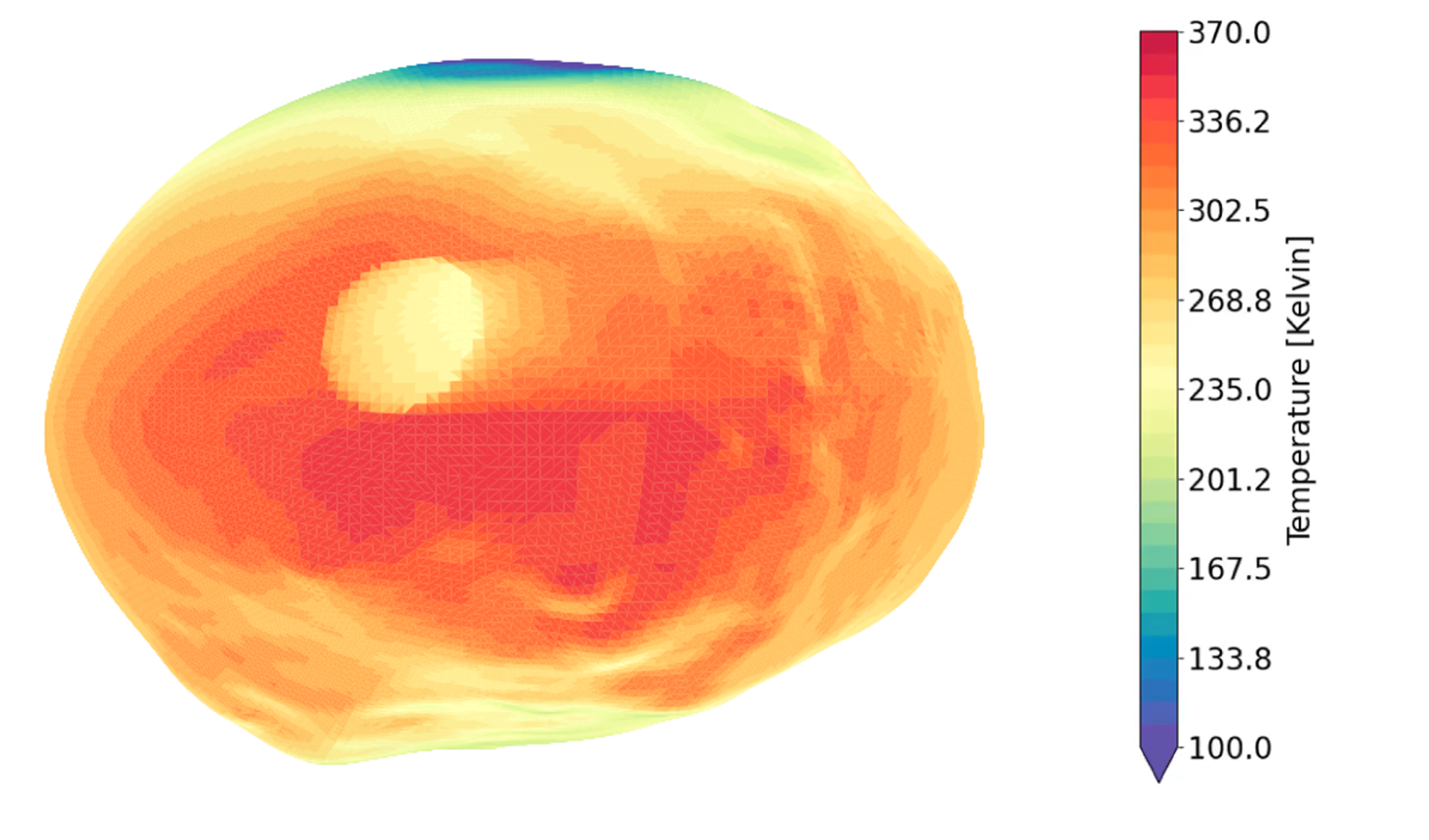}
            \caption{Prediction of the presence of a thermal wake on the surface of 65803 Didymos during the Hera mission operational phase using a thermal inertia of $\Gamma = 320 \pm 70\,\tiu$ \citet{2024PSJ.....5...66R}.}
            \label{fig:didymosWake}
        \end{figure}

        For example, when modelling Didymos we can make use of a low-resolution shape model for Dimorphos to calculate the effect of eclipse and occultations.
        We can also neglect the calculation of viewfactors as self-heating effects from Dimorphos onto Didymos will be negligible.
        If we simulate the Didymos system at an early point during the Hera mission timeline (2027-01-01) using the thermal inertia ($\Gamma = 320 \pm 70\,\tiu$) derived by \citet{2024PSJ.....5...66R}, we predict that a thermal wake will be observable on the surface of Didymos as seen in Figure~\ref{fig:didymosWake}.

        When modelling Dimorphos we would invert the resolution of the shape models and use a low-resolution shape model of Didymos to calculate eclipses and occultations.
        This time, however, we can rotate Didymos at the same rate as Dimorphos' orbit and therefore use a single-object model with two disjoint components.
        This enables viewfactor calculations to be calculated only once, rather than calculated at each step in the rotation/orbit.
        Rotating Didymos slower would change its surface temperature distribution, and therefore impact the degree of self-heating from Didymos onto Dimorphos.
        The interplay between thermal inertia and rotation on the surface temperature of an asteroid can be described by the dimensionless thermal parameter, $\Theta = \frac{\Gamma\sqrt{\omega}}{\varepsilon\sigma T_{SS}^4}$, where $\Gamma$ is the thermal inertia, $\omega$ is the rotation rate of the asteroid, $\varepsilon$ is the emissivity, $\sigma$ is the Stefan-Boltzmann constant, and $T_{SS}$ is the sub-solar temperature.
        If we reduce the rotation rate of the asteroid in this case, then we can increase the thermal inertia accordingly to keep the dimensionless thermal parameter constant, and therefore produce a similar temperature distribution on the surface for the sake of self-heating calculations from Didymos to Dimorphos.

        These assumptions may enable us to use the ATPM and the binary adaptations described in this work to model the TIRI data from ESA's Hera mission while significantly reducing computational time required.
        Precise details on how to model the system will need to be adapted before arrival at the Didymos system, to account for the fact that the ratio of the primary rotation to secondary orbit periods is not exactly 5:1, in order to prevent mis-timed eclipses and occultations in the model biasing our results.
        The inclusion of potential tumbling or librating states caused by the impact of the DART spacecraft into Dimorphos \citep{2023PSJ.....4..141M} may also need to be included to accurately model the system, and how to include this effect remains subject to future work.

    \subsection{Other Applications}

        NASA's Lucy mission to the Trojan asteroids \citep{2021PSJ.....2..171L} will visit a variety of binary asteroid systems such as (15094) Polymele, (21900) Orus, and (617) Patroclus, and the binary ATPM could be used to model their thermophysical properties from thermal emission spectra from the L'TES instrument \citep{2024SSRv..220....1C}.
        The recent engineering test flyby of the asteroid 152830 Dinkinesh by the Lucy spacecraft found that this asteroid also has a companion, Selam.
        The ATPM and its binary adaptations could be applied to the L'TES data collected during this flyby, if they are of sufficient quality and the orbit of the secondary can be determined from spacecraft or ground-based data.

        Other binary systems are also very likely to have IR data collected by the \textit{WISE} telescope over the years, and so a systematic search for data that should be reprocessed to account for mutual event features in their thermal-IR light curves would be wise to improve our corpus of binary asteroid thermophysical properties.

    \subsection{Future Development}\label{sec:future}

        The model described in this paper is designed to handle these fully and semi-synchronous systems, but can not handle systems where there is no integer ratio between the primary rotation and secondary orbital periods.
        Naturally there will always be a ratio of integers that could represent the two periods, but if this ratio requires more than 10 or so rotations of one component the computation will be excessive.
        A significant re-write of the ATPM would be required to include such systems, as the code would need to run significantly faster and output much less intermediate data.
        To model these systems without significant overheads, a method of using long-term quasi-stability to converge the temperatures is likely needed.
        This approach would involve modelling the asteroid over an entire orbit or two to ensure model temperatures have converged to a `quasi-stable' state at the point in its orbit at which we wish to model it.
        This is beyond the scope of the work in this paper, but may be implemented in a future re-write of the model.

\section{Summary}

    In this work we have outlined adaptations made to the Advanced Thermophysical Model \citep[ATPM;][]{2011MNRAS.415.2042R,2012MNRAS.423..367R,2013MNRAS.433..603R} to enable a more thorough interpretation of thermal-IR light curves of binary asteroid systems.
    Limitations on targets that can be studied with this model have been outlined, i.e. we can only model systems with synchronous or semi-synchronous primary spin to secondary orbit periods that are not experiencing significant tumbling or libration.
    We model \textit{WISE} data of the synchronous binary (3905) Doppler to determine a thermal inertia of $\Gamma = 114 \pm 31\,\tiu$ and a roughness fraction of $0.73 \pm 0.22$ (corresponding to an RMS slope of $41 \pm 6\si\degree$).
    We find tentative evidence that a pole orientation that maximises occultations and removes eclipses from the viewing geometry may explain the data better, although the reduced chi-square values suggest this may be over-fitting (which is likely given the heavy dependence on a single data point).
    Thermal-IR light curves from \textit{WISE} and \textit{NEOWISE} of the approximately semi-synchronous binary (175706) 1996 FG3 are modelled using the new binary TPM, deriving a thermal inertia of $\Gamma = 142 \pm 6\,\tiu$, and a roughness fraction of $f_R = 0.31 \pm 0.06$ (corresponding to an RMS slope of $27 \pm 3\si\degree$).
    
    The sensitivity of binary thermal-IR light curves has been discussed, showing strong modulations to the light curves for high values of the thermal inertia due to the presence of `thermal wakes' in semi-synchronous systems that weaken gradually over time.
    The thermal wake is deemed to be most important when modelling shorter wavelength data due to the increased sensitivity to high temperatures, and the wavelength of observation is also shown to modify the morphology of eclipse events in these thermal-IR light curves.
    The application of binary thermophysical models to missions such as ESA's Hera mission and NASA's Lucy mission has been discussed, with ample opportunity for binary systems to be modelled in detail over the coming decade, providing a greater understanding of this important population.

\section*{Acknowledgements}

SLJ and BR are funded by the Science and Technology Facilities Council under grant ST/X001180/1.
This publication uses data products from NEOWISE, a project of the Jet Propulsion Laboratory/California Institute of Technology, funded by the Planetary Science Division of NASA.
We made use of the NASA/IPAC Infrared Science Archive, which is operated by the Jet Propulsion Laboratory/California Institute of Technology under a contract with NASA.

\section*{Data Availability}

The WISE data used in this work are all available from the WISE/NEOWISE catalogue at \url{https://irsa.ipac.caltech.edu/cgi-bin/Gator/nph-scan?mission=irsa&submit=Select&projshort=WISE}.
The Binary ATPM code used in this work is available at \url{https://github.com/SamuelLeeJackson/BinaryTPM}.

\bibliographystyle{mnras}
\bibliography{references}

\appendix
\section{Supplementary Figures}

    \begin{figure*}
        \centering
        \includegraphics[width=\linewidth]{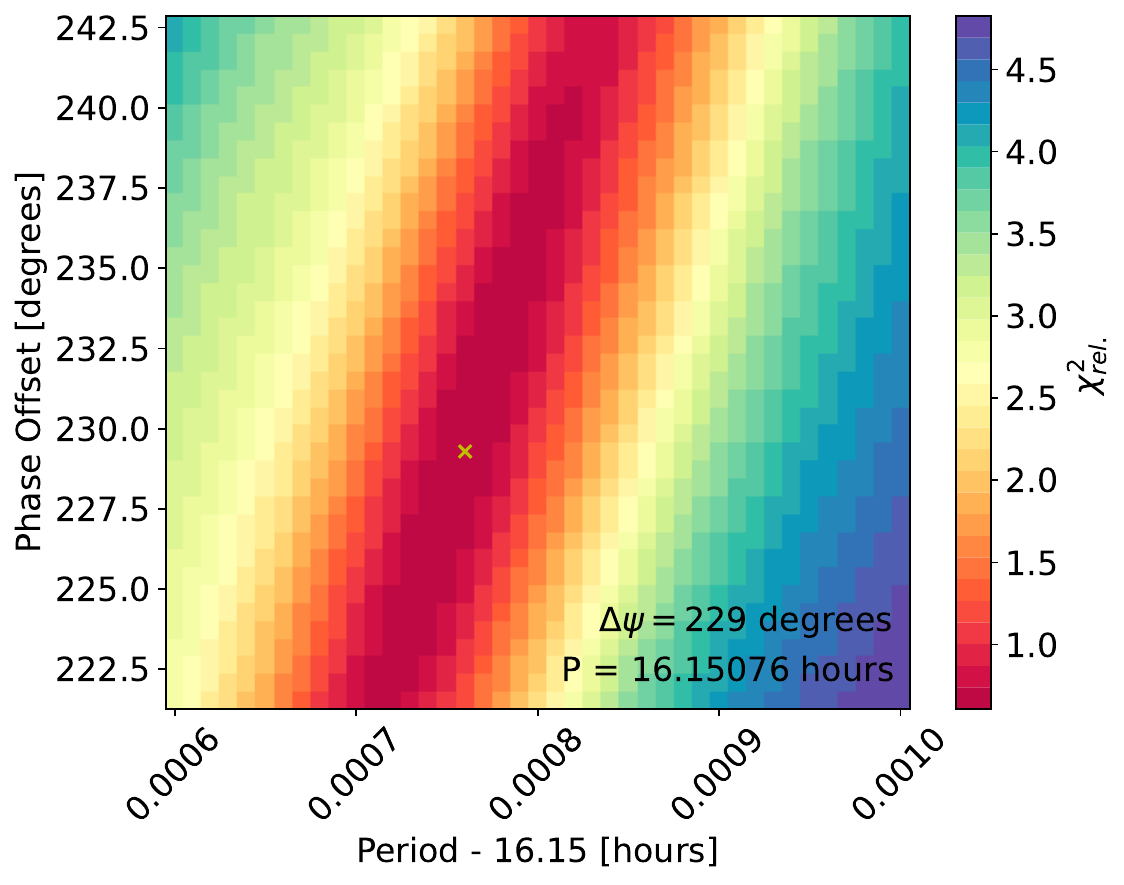}
        \caption{Chi-squared plane evaluating the phase offset and orbital period needed to best describe optical data from \citet{2015Icar..245...56S} using our constructed model of the binary system. This phasing to the optical data is then used to ensure that the thermophysical model fits to the \textit{WISE} data are phased appropriately.}
        \label{fig:FG3phasing}
    \end{figure*}
    
    \begin{figure*}
        \centering
        \includegraphics[width=.95\textwidth]{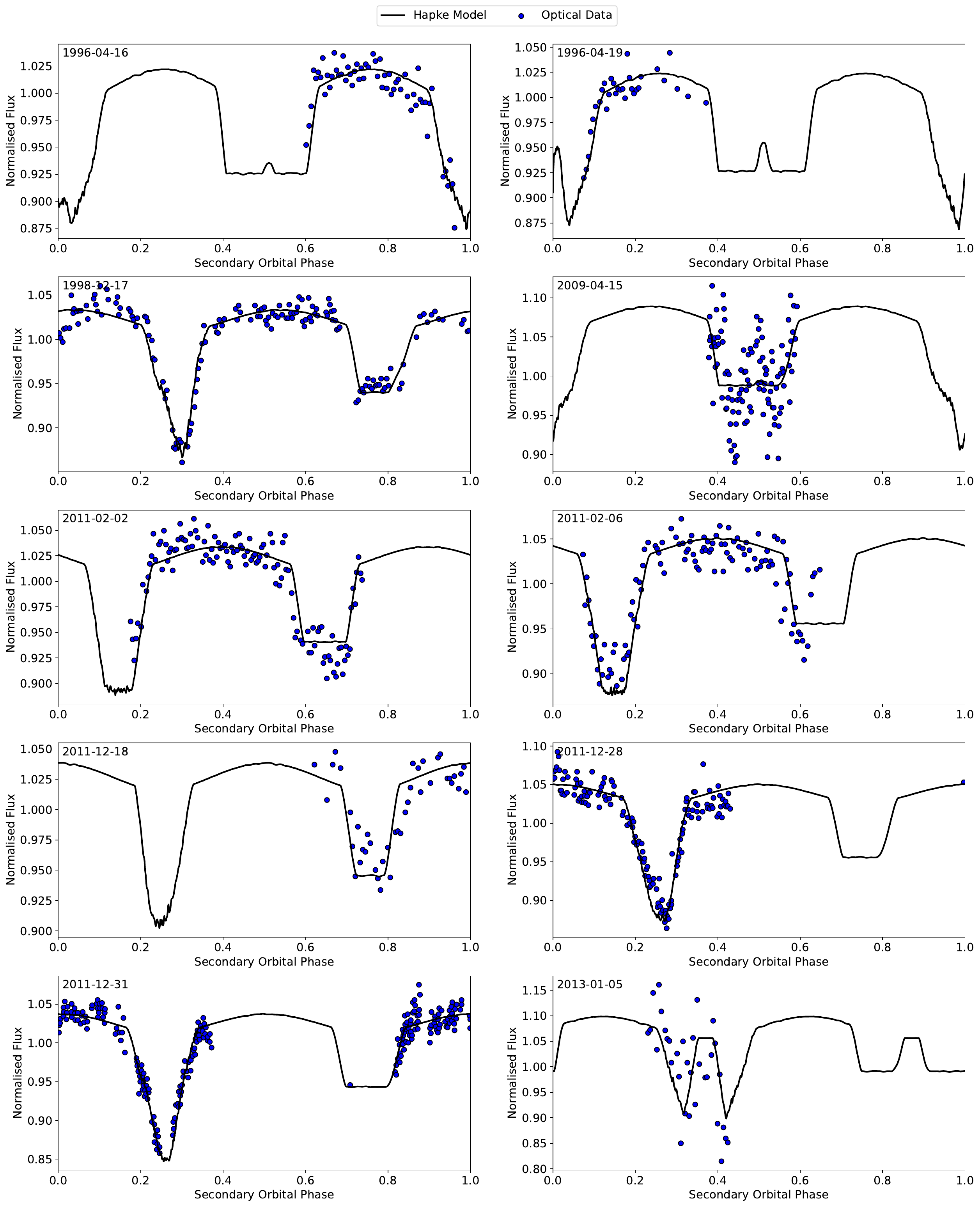}
        \caption{Lambert/Lommel-Seeliger model predicted optical fluxes using the derived model phase offset and orbital period, compared to the optical photometry digitised from \citet{2015Icar..245...56S}. It can be seen that the model reproduces the mutual events in the optical data, and therefore is considered to be adequately in-phase for use in the thermophysical modelling process.}
        \label{fig:optical_fits}
    \end{figure*}

    \begin{figure*}
        \centering
        \includegraphics[width=.95\textwidth]{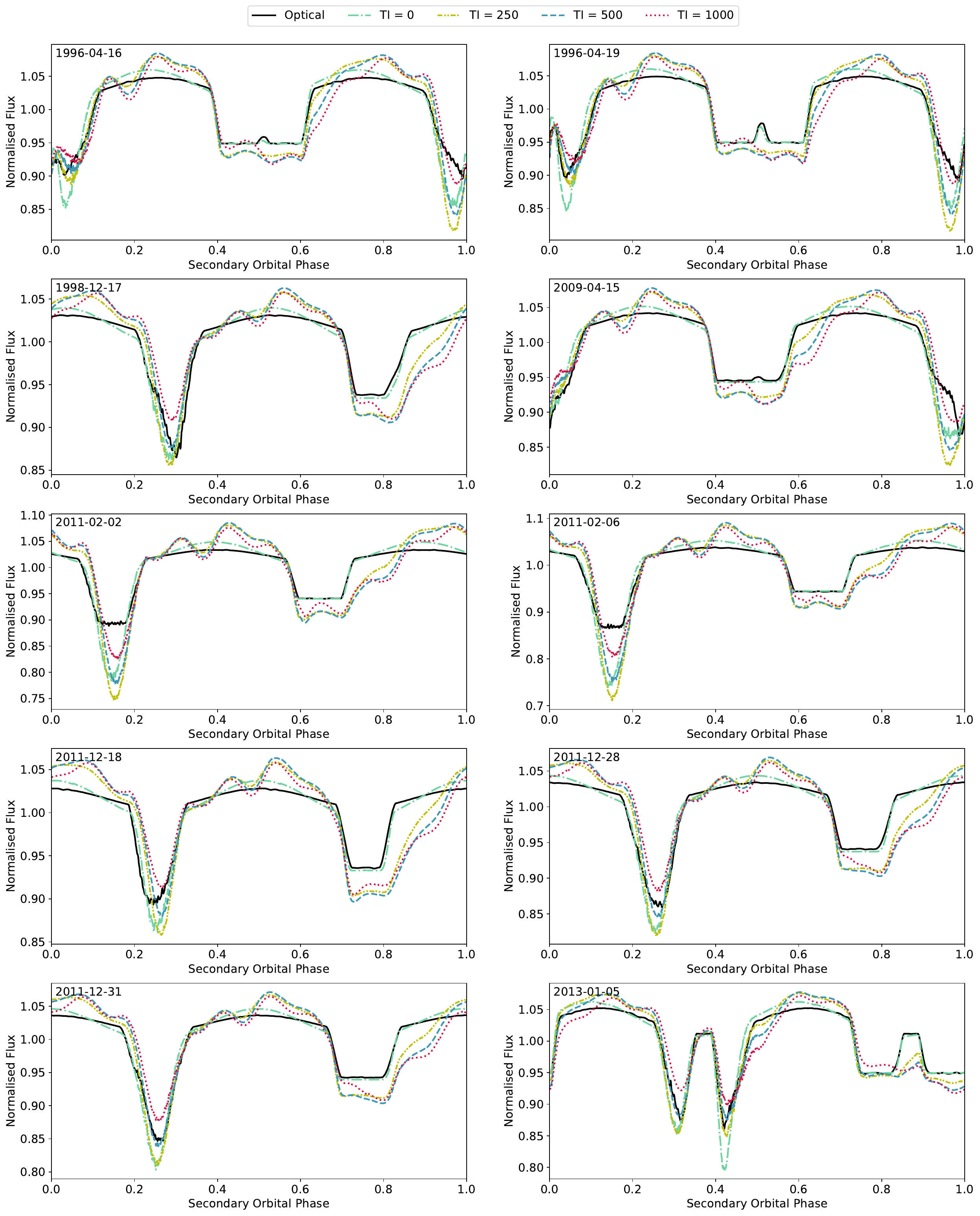}
        \caption{Visualisations of the model-produced optical and thermal light curves for each observing geometry from \citet{2015Icar..245...56S}. The optical model (black solid line) is plotted for comparison against the predicted smooth-surface thermal light curves in the WISE W2 band is shown for thermal inertias of 0 (green dash-dotted line), 250 (yellow dash-dot-dotted line), 500 (blue dashed line), and 1000 (red dotted line) $\tiu$.}
        \label{fig:ti_study}
    \end{figure*}
    \begin{figure*}
        \centering
        \includegraphics[width=.95\textwidth]{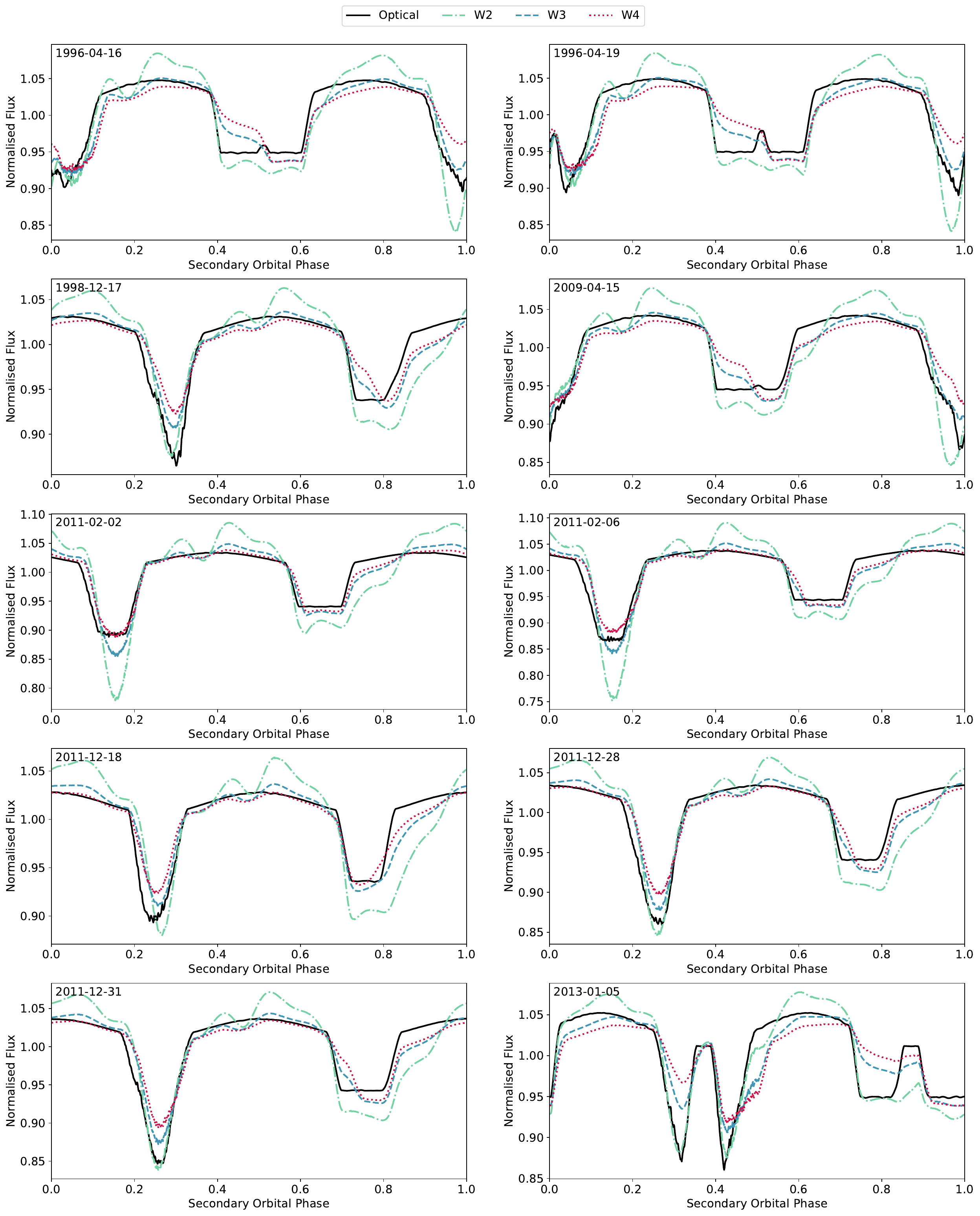}
        \caption{Visualisations of the model-produced optical and thermal light curves for each observing geometry from \citet{2015Icar..245...56S}. The optical model (block solid line) is plotted for comparison against the predicted smooth-surface thermal light curves with a thermal inertia of $500 \tiu$ is shown for the W2 (green dash-dotted line), W3 (blue dashed line), and W4 (red dotted line) WISE bands.}
        \label{fig:wavelength_study}
    \end{figure*}
    \begin{figure*}
        \centering
        \includegraphics[width=.95\textwidth]{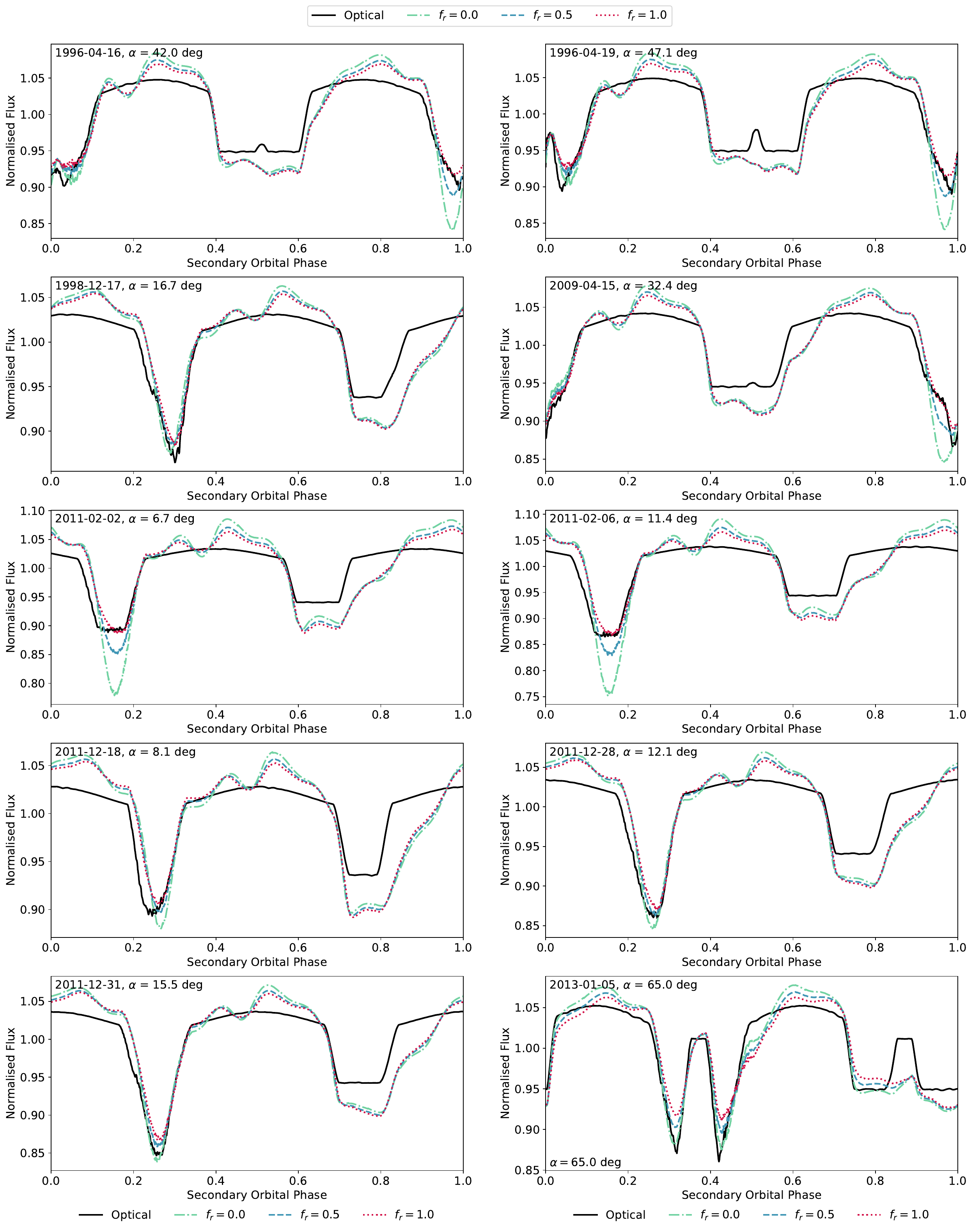}
        \caption{Visualisations of the model-produced optical and thermal light curves for each observing geometry from \citet{2015Icar..245...56S}. The optical model (black solid line) is plotted for comparison against the predicted thermal light curves for a thermal inertia of $500 \tiu$ in the WISE W2 band for roughness fractions of 0.0 (green dash-dotted line), 0.5 (blue dashed line), and 1.0 (red dotted line).}
        \label{fig:roughness_study}
    \end{figure*}

\bsp
\label{lastpage}
\end{document}